\documentclass[sn-nature]{sn-jnl}% Style for submissions to Nature Portfolio journals
\usepackage{multirow}%
\usepackage{amsmath,amssymb,amsfonts}%
\usepackage{amsthm}%
\usepackage{mathrsfs}% 
\usepackage[title]{appendix}%
\usepackage{xcolor}%
\usepackage{textcomp}%
\usepackage{manyfoot}%
\usepackage{booktabs}%
\usepackage{listings}%

\usepackage{caption}

\usepackage{ulem}
\renewcommand{\emph}[1]{\textit{#1}}

\usepackage{xspace}

\usepackage{comment}
\newcommand{\Fermi}{\emph{Fermi}\xspace}
\newcommand{\fermi}{\emph{Fermi}\xspace}

\newcommand{\Swift}{\emph{Swift}\xspace}

\definecolor{blazeorange}{rgb}{1.0, 0.4, 0.0}
\definecolor{seagreen}{rgb}{0.18, 0.55, 0.34}
\definecolor{rufous}{rgb}{0.66, 0.11, 0.03}
\definecolor{royalfuchsia}{rgb}{0.79, 0.17, 0.57}
\definecolor{scarlet}{rgb}{1.0, 0.13, 0.0}
\definecolor{royalpurple}{rgb}{0.47, 0.32, 0.66}
\definecolor{darkblue}{rgb}{0, 0, 0.66}
\newcommand\apj{Astrophysical Journal}% Astrophysical Journal ++
\newcommand\apjl{Astrophysical Journal Letters} % Astrophysical Journal, Letters
\newcommand\apjs{Astrophysical Journal Supplement}% Astrophysical Journal Supplements
\newcommand\nat{Nature}%  % Nature 
\newcommand\mnras{Mon. Not. R. Astron. Soc.}%Monthly Notices of the Royal Astronomical Society}%   % Monthly Notices of the RAS
\newcommand\aap{Astronomy and Astrophysics}%     % Astronomy and Astrophysics 
\newcommand\physrep{Physics Reports}%    % Physics Reports 

\theoremstyle{thmstyleone}%
\theoremstyle{thmstyletwo}%

\theoremstyle{thmstylethree}%

\begin{document}

\title{Gamma rays from a reverse shock with turbulent magnetic fields in GRB~180720B}
%\title{Gamma rays and optical polarization unveil the magnetic fields in the shocks of GRB~180720B}

%\author[0]{The \fermi LAT and Kanata collaborations}
%\affil[0]{Category II paper}

\author[1,*]{Makoto Arimoto}
%\author[1,*]{Arimoto Makoto}
\affil[*]{Contact Author}
\affil[1]{Institute of Science and Engineering, Kanazawa University, Kakuma, Kanazawa, Ishikawa 920-1192, Japan}

\author[2]{Katsuaki Asano}
%\author[2]{Asano Katsuaki}
\affil[2]{Institute for Cosmic Ray Research, The University of Tokyo, 5-1-5 Kashiwanoha, Kashiwa, Chiba 277-8582, Japan}

\author[3,4]{Koji S. Kawabata}
%\author[3]{Kawabata S. Koji}
\affil[3]{Hiroshima Astrophysical Science Center, Hiroshima University, 1-3-1 Kagamiyama, Higashi-Hiroshima, Hiroshima 739-8526, Japan}
\affil[4]{Department of Physical Science, Hiroshima University, 1-3-1 Kagamiyama, Higashi-Hiroshima, Hiroshima 739-8526, Japan}

%\begin{comment}

\author[5]{Kenji Toma}
%\author[4]{Toma Kenji}
\affil[5]{Frontier Research Institute for Interdisciplinary Sciences, Tohoku University, Sendai 980-8578, Japan}

\author[6,7]{Ramandeep Gill}
%\author[5,7]{Gill Ramandeep}
%\affil[6]{Instituto de Radioastronom\UTF{00ED}a y Astrof\UTF{00ED}sica, Universidad Nacional Aut\UTF{00F3}noma de M\UTF{00E9}xico, Antigua Carretera a P\UTF{00E1}tzcuaro \# 8701, 
%Ex-Hda. San Jos\UTF{00E9} de la Huerta, Morelia, Michoac\UTF{00E1}n, M\UTF{00E9}xico C.P. 58089}
\affil[6]{Instituto de Radioastronom\'ia y Astrof\'isica, Universidad Nacional Aut\'onoma de M\'exico, Antigua Carretera a P\'atzcuaro \# 8701, 
Ex-Hda. San Jos\'e de la Huerta, Morelia, Michoac\'an, M\'exico C.P. 58089}

\affil[7]{Astrophysics Research Center of the Open university (ARCO), The Open University of Israel, P.O Box 808, Ra'anana 4353701, Israel}

\author[8,7,9]{Jonathan Granot}
%\author[6,7,8]{Granot Jonathan}
\affil[8]{Department of Natural Sciences, The Open University of Israel, P.O Box 808, Ra'anana 4353701, Israel}
\affil[9]{Department of Physics, The George Washington University, Washington, DC 20052, USA}

\author[4,3]{Masanori Ohno}
%\author[3]{Ohno Masanori}
\author[1]{Shuta Takahashi}
%\author[1]{Takahashi Shuta}
\author[1]{Naoki Ogino}
%\author[1]{Ogino Naoki}
\author[1]{Hatsune Goto}
%\author[1]{Goto Hatsune}
\author[4]{Kengo Nakamura}
%\author[3]{Nakamura Kengo}
\author[3,4]{Tatsuya Nakaoka}
%\author[3]{Nakaoka Tatsuya}
\author[4]{Kengo Takagi}
%\author[3]{Takagi Kengo }
\author[10,4]{Miho Kawabata}
%\author[9]{Kawabata Miho}
\affil[10]{
Nishi-Harima Astronomical Observatory, Center for Astronomy, University of Hyogo, 407-2 Nishigaichi, Sayo, Hyogo 679-5313, Japan}
\author[11,3]{Masayuki Yamanaka}
%\author[9]{Yamanaka Masayuki}
%\affil[9]{Department of Astronomy, Kyoto University, Kitashirakawa-Oiwake-cho, Sakyo-ku, Kyoto 606-8502, Japan}
\affil[11]{
Amanogawa Galaxy Astronomy Research Center, Kagoshima University, 1-21-35 Korimoto, Kagoshima, Kagoshima 890-0065,Japan}

\author[12,3]{Mahito Sasada}
%\author[10]{Sasada Mahito}
\affil[12]{
Department of Physics, Tokyo Institute of Technology, 2-12-1 Ookayama, Meguro-ku, Tokyo 152-8551, Japan}
\author[13,9]{Soebur Razzaque}
%\author[11,8]{Razzaque Soebur}
\affil[13]{Centre for Astro-Particle Physics (CAPP) and Department of Physics, University of Johannesburg, PO Box 524, Auckland Park 2006, South Africa}
%\end{comment}

%\keywords{Keyword1, Keyword2, Keyword3}

%\begin{abstract}
%\textbf{
\abstract{
Gamma-ray bursts (GRBs) are the most electromagnetically luminous cosmic explosions. They are powered by collimated streams of
plasma (jets) ejected by a newborn stellar-mass black hole or neutron star at relativistic velocities (near the speed of light).
 Their short-lived  (typically tens of seconds) prompt $\gamma$-ray emission from within the ejecta is followed by long-lived multi-wavelength afterglow emission from the ultra-relativistic forward shock. This shock is driven into the circumburst medium by the GRB ejecta that are in turn decelerated by a mildly-relativistic reverse shock. 
 Forward shock emission was recently detected up to teraelectronvolt-energy $\gamma$-rays, and such very-high-energy emission was also predicted from the reverse shock. 
 Here we report the detection of optical and gigaelectronvolt-energy $\gamma$-ray emission from GRB~180720B during the first few hundred seconds, which is explained by synchrotron and inverse-Compton emission from the reverse shock propagating into the  ejecta, implying a low-magnetization ejecta. Our optical measurements show a clear transition from the reverse shock to the forward shock driven into the circumburst medium, accompanied by a 90-degree change in the mean polarization angle and fluctuations in the polarization degree and angle.
 This indicates turbulence with large-scale toroidal and radially-stretched magnetic field structures in the reverse and forward shocks, respectively, which tightly couple to the physics of relativistic shocks and GRB jets -- launching, composition, dissipation and particle acceleration.
}
%\end{abstract}

%\begin{document}
%\flushbottom
\maketitle

\begin{figure*}[h!]   
\begin{center}
  \hspace*{0em} % move the table to left
\includegraphics[width=1.1\textwidth, viewport=0 0 610 642, angle=0]{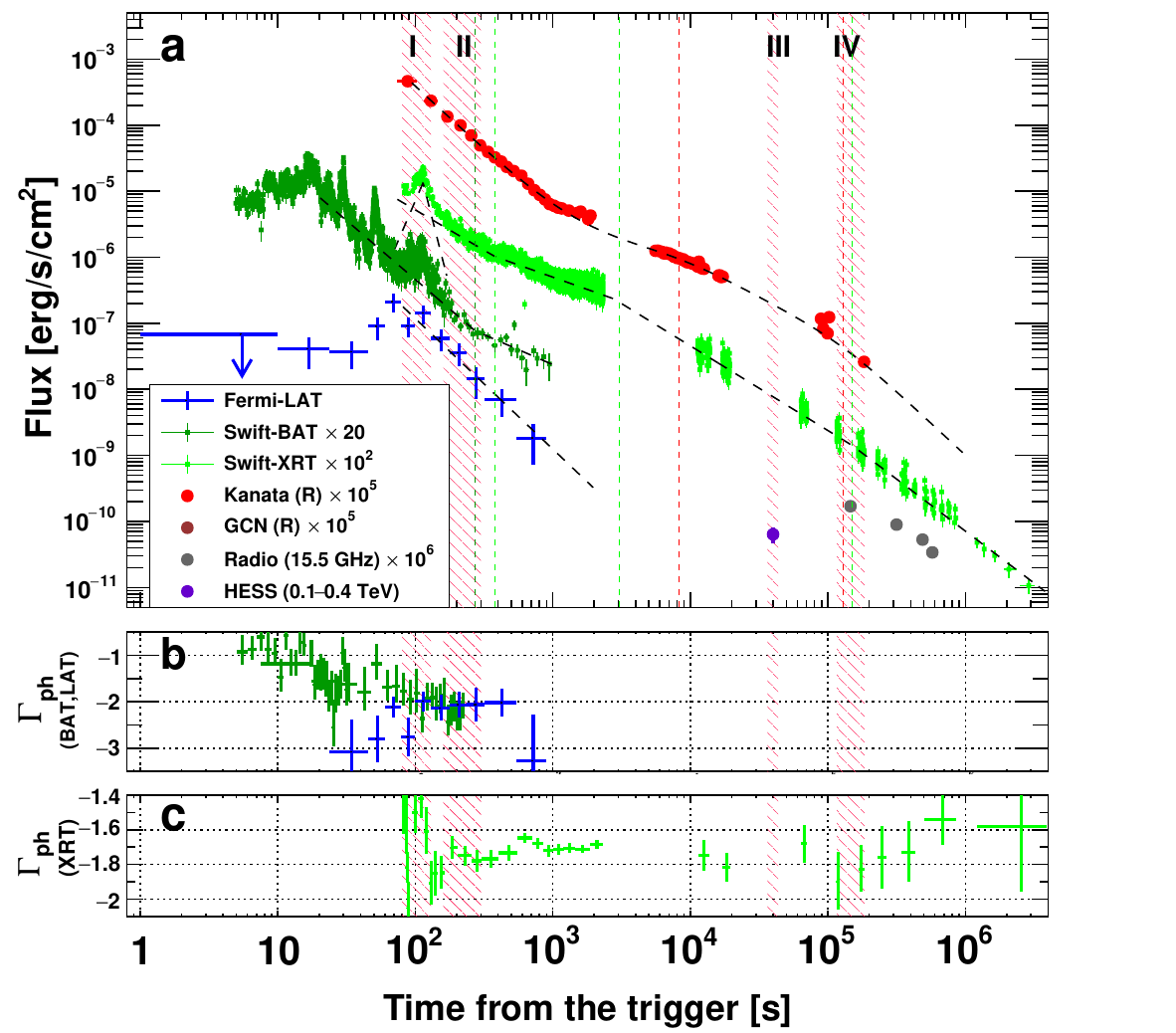}
\caption{\textbf{Lightcurves from the radio to TeV bands of GRB~180720B.}
({\bf a}) The unabsorbed lightcurves of GRB~180720B  from \Fermi-LAT (0.1--1 GeV), \Swift-BAT (15--150 keV), \Swift-XRT (0.3--10 keV), Kanata and other telescopes denoted by GCN ($\sim$eV), AMI-LA (15.5 GHz), and HESS (0.1--0.4 TeV). The black dashed lines represent the best-fitting power-law functions with breaks and the vertical dashed lines represent the corresponding break times.
({\bf b}) The observed photon indices observed by  \Swift-BAT (dark green) and \Fermi-LAT (blue).
({\bf c}) The observed photon indices observed by \Swift-XRT (green). All error bars correspond to the 1-$\sigma$ confidence region.
}
\label{fig:LC} 
\end{center}   
\end{figure*}

On 20 July 2018, the \Fermi Gamma-ray Burst Monitor (GBM) and {\it Swift} Burst Alert Telescope (BAT) triggered and localized GRB~180720B at a redshift of $z=0.654$\cite{2018GCN.22996....1V} (a luminosity distance of 4.0 Gpc), with a time-integrated isotropic equivalent energy of $E_{\rm iso}=5\times10^{53}\;$erg (1$\,$--$\,$10$^4\;$keV) and a duration of $\sim60\;$s (see Supplementary Methods).
It was followed by multi-wavelength observations from radio to TeV energies: radio, optical, X-ray, GeV, TeV\cite{2019Natur.575..464A}, where \Fermi Large Area Telescope (LAT) covers the GeV band (see Methods). The observed lightcurves in different energy bands are shown in Fig.~\ref{fig:LC}. 

\begin{figure*}[ht!]
    \centering
    \hspace*{-7em} % move the table to left
 \includegraphics[width=1.0\textwidth, viewport=0 0 610 642]{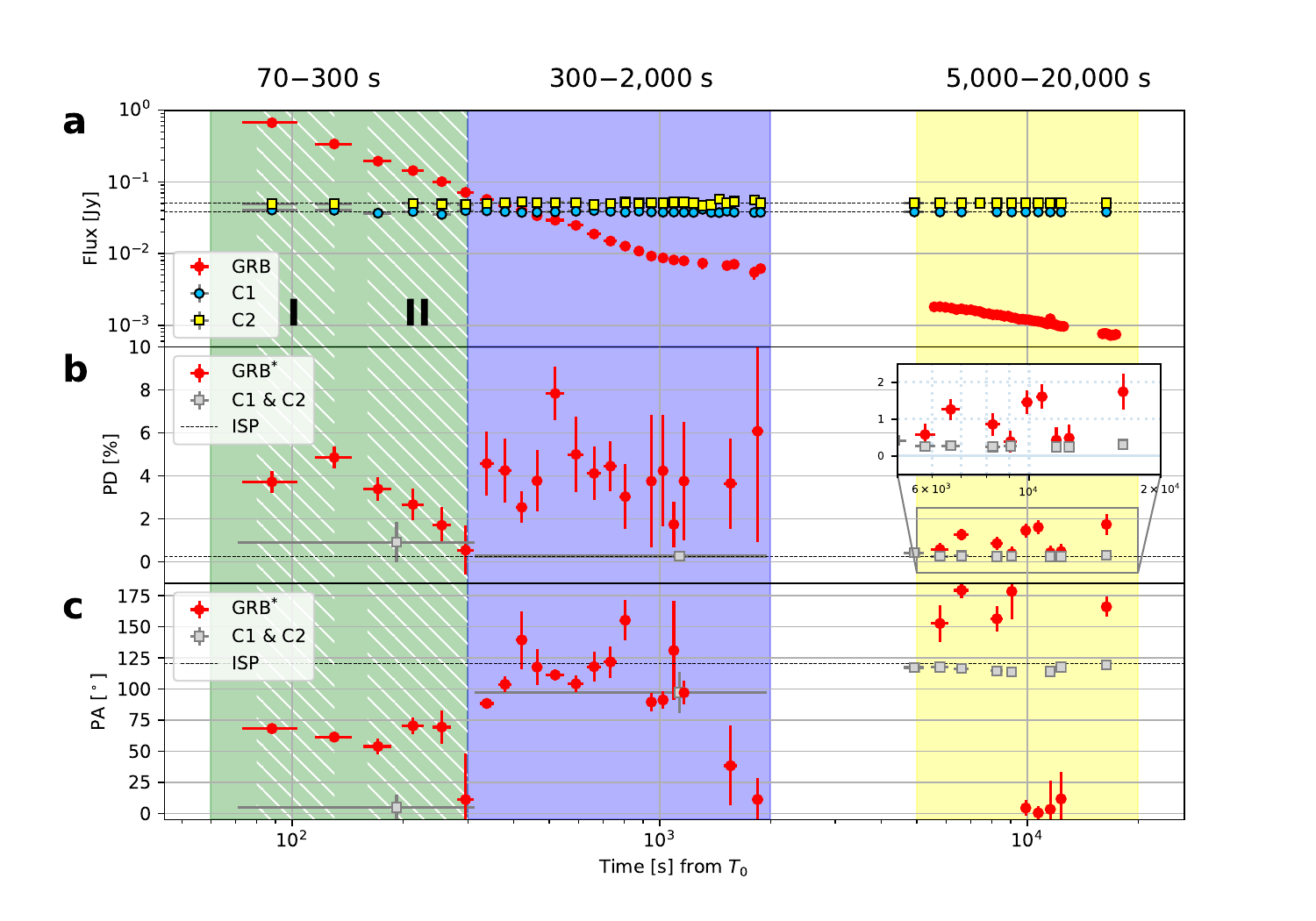}
  \caption{\textbf{Optical lightcurve, polarization degree and angle of GRB~180720B and nearby stars.}
  ({\bf a}) The optical lightcurves of GRB~180720B and the nearby stars (C1 and C2) observed by HOWPol and HONIR implemented on the Kanata telescope. 
  ({\bf b}) The polarization degrees (PDs) of GRB~180720B and the average of the nearby stars (C1 \& C2). The inset figure shows a zoomed plot of the observed PD points of this GRB, C1 and C2. 
  ({\bf c}) The polarization angles (PAs) of GRB~180720B and the average of the nearby stars (C1 \& C2). A ``$*$'' symbol indicates the intrinsic GRB polarization after subtraction of the interstellar polarization (ISP). HOWPol and HONIR covered $T_0$ + 70 -- 2,000 s and $T_0$ + 5,000 -- 20,000 s, respectively.  The hatched areas represent the time intervals (I and II) shown in Fig. \ref{fig:LC}. All error bars correspond to the 1-$\sigma$ confidence region.
}
\label{fig:QU} 
\end{figure*}

Following the alert of the GRB position by \Swift, the Kanata 1.5-m telescope performed followup 
observations.\cite{2018GCN.22977....1S} Equipped with optical polarimetry instruments (HOWPol and HONIR) it detected 
bright optical emission $\sim$100 s after the burst trigger (GRB trigger time represented as $T_0$), while the \Fermi-LAT also detected 
bright GeV emission peaking at $T_0$ + $\sim100\;$s. 
In the early phase ($T_0+\, 100\;$--$\,1000\;$s), the optical ($F_{\rm opt}\propto{}t^{\alpha_{\rm opt}}$) 
and GeV ($F_{\rm GeV}\propto{}t^{\alpha_{\rm GeV}}$) fluxes declined with temporal indexes of 
$\alpha_{\rm opt}=-1.94\pm0.08$ and $\alpha_{\rm GeV}=-1.91\pm0.31$, respectively, showing a similar trend.
Both values are much steeper than the typical temporal index of GRB afterglows,\cite{1998ApJ...497L..17S, 2019ApJ...878...52A} 
indicating rapidly fading emission originating from a reverse shock.   
During the subsequent time, the optical temporal index 
became $\alpha_{\rm opt}=-1.10\pm0.02$ (See Methods), which is a typical index of emission coming from the  forward shock. 
Our optical polarization measurements by HOWPol and HONIR during $T_0+70-20,000\;$s cover both the reverse and forward shock dominated phases (Fig.~\ref{fig:QU}).
They reveal that the polarization degree (PD) and polarization angle (PA) were changing gradually (PD$\;\lesssim\,1\,$--$\;8$\% and PA$\;\sim50^\circ$ to $\sim150^\circ$) during the initial 1000-s interval after the burst. 
In the late phase ($t_{\rm obs} > T_{\rm 0} + 5000$ s),  almost constant PD and PA ($\sim$1\% with $\sim$160$^\circ$) were detected (See Supplementary Methods for the analysis details). Detection of optical polarization from the reverse and forward shocks in a single GRB is unprecedented\cite{2013Natur.504..119M},
and can be a powerful probe of the structure and origin of magnetic fields in the shocked regions.

\begin{figure*}[ht!]   
    \centering
    \hspace*{-6em} % move the table to left
    \includegraphics[width=1.0\textwidth,viewport=0 0 610 642]{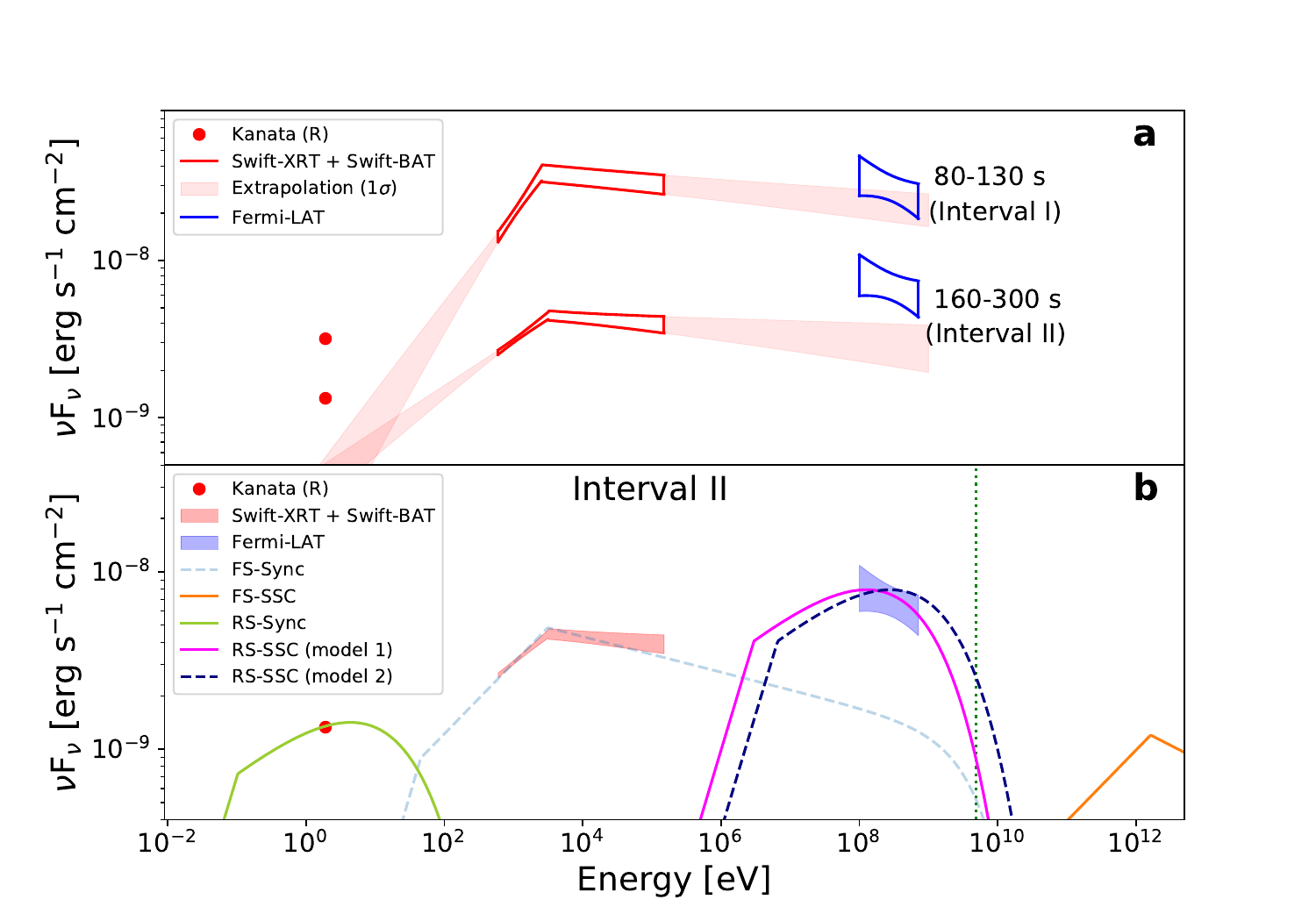}
\caption{\textbf{Spectral energy distributions from $T_0$ + 80 s to 300 s and theoretical modeling at time interval II.}
The red and blue areas correspond to the 1-$\sigma$ confidence regions from the best-fit functions (i.e., a broken power-law function and a simple power-law function, respectively) for the \Swift-XRT + BAT  and \Fermi-LAT ranges, respectively.  The red points represent the optical flux observed by the Kanata telescope.
({\bf a}) The pale red shaded regions correspond to the 1-$\sigma$ region extrapolated from the \Swift-XRT and BAT range. 
({\bf b})  The solid yellow green line represents the synchrotron component from the reverse shock. The solid magenta  and the dashed navy lines represent the SSC components with ``model 1'' and ``model 2'',  respectively (see Methods for more details). The dashed light blue and the solid orange  lines represent the synchrotron and SSC components of the forward shock, respectively. The  green vertical dotted line corresponds to the highest energy photon of 5 GeV (see Supplementary Methods).
}
\label{fig:SED_multi}   
\end{figure*}

First, to better understand the rapid fading in the early phase, we extracted wide-band (optical to GeV) spectral energy distributions (SEDs) from $T_0$ + 80 s to 300 s (Fig. \ref{fig:SED_multi}).  
Specifically in the time interval $T_0+160-300\;$s
(Interval II), the optical and GeV components are distinctly higher than the extrapolations from the X-ray component likely originating from the forward shock (See Supplementary Methods for the significance of the GeV excess). Thus, to explain the optical and GeV excesses, an additional component such as the reverse-shock component is needed. 
Note that in the time interval $T_0+80-130\;$s
(Interval I), the flux contribution from a bright X-ray flare at $T_0\,+\,\sim\!100\;$s is significant and the X-ray flare likely comes from a
different emission site as indicated 
by the short variability timescales (see Methods for a discussion of the X-ray flare).

The GeV onset timescale of $\sim$100 s roughly corresponds to the time required 
for the reverse shock to cross the ejecta shell, and since the prompt emission lasts $\sim$60 s, this implies that
the reverse shock is mildly relativistic\cite{1995ApJ...455L.143S}. The bright optical emission can be explained by 
synchrotron emission from the reverse shock in the slow cooling regime\cite{2000ApJ...545..807K}. 
However, after the reverse shock crosses the ejecta shell there is no injection of freshly accelerated power-law electrons 
into the shocked shell. As a result, synchrotron emission above the synchrotron cooling frequency sharply drops, where the typical cooling frequency reaches the X-ray band at most\cite{2000ApJ...545..807K}. 
Thus, the observed $\gamma$-rays cannot be powered by synchrotron emission, and are potentially produced by synchrotron self-Compton emission (SSC) in the reverse shock,\cite{2001ApJ...546L..33W,2001ApJ...556.1010W,2001ApJ...559..110Z} with seed optical photons inverse-Compton scattered to GeV energies by the same shock-accelerated electrons emitting the optical photons. 
The theoretical temporal index of the SSC emission from the reverse shock is almost similar to that of the synchrotron emission (see Methods).
 Several works have already reported possible SSC emission from the reverse shock\cite{2016ApJ...818..190F,2017ApJ...848...15F,2019ApJ...879L..26F,2020ApJ...905..112F, 2023ApJ...947L..14Z}, where they relied on only a few simultaneous optical observations to characterize both the reverse and forward shocks distinctly for the first few thousand seconds.

Because the intensity of the SSC emission depends on the fraction of the internal energy 
held by the electrons ($\epsilon_{e, r}$) and the magnetic field ($\epsilon_{B, {\rm r}}$) 
in the reverse shock,\cite{1998ApJ...497L..17S} 
the observed $\gamma$-ray emission can constrain these two microphysical parameters. 
To reproduce the observed SSC-to-synchrotron flux ratio ($Y \sim$ 6) 
of the SED in Interval II, a small value of $\epsilon_{B, {\rm r}}$ is required ($\epsilon_{B, {\rm r}} \sim$ 10$^{-4}$--10$^{-3}$, See Methods and Extended Data Table \ref{table:parameters_rs_fs}).
The theoretical models fit well the observed optical and GeV fluxes, as shown in Fig.~\ref{fig:SED_multi}, Extended Data Figs. \ref{fig:LC-model} and \ref{fig:EATS-model}. 
Due to the soft spectrum in the GeV band observed by \Fermi-LAT, our modeling suggests a low value of the maximum electron energy
in the reverse shock region. The particle acceleration efficiency would be suppressed in the reverse shock region compared to that in the forward shock (See Methods).
However, because the uncertainty of the \Fermi-LAT spectrum is relatively large, future simultaneous 
observations by TeV Cherenkov telescopes, such as MAGIC and 
the Cherenkov Telescope Array\cite{2019scta.book.....C}, 
during the early phase of the GRB afterglow  will give a stringent limit on this process.

In the late phase ($t_{\rm obs}$ $\gtrsim$ $T_0$ + 5000 s), emission from the reverse shock does not contribute to the observed fluxes due to its steep temporal decline,
and forward shock emission is dominant. 
Our analytical model with synchrotron and SSC emission in the forward shock matches the observed spectrum from the optical to VHE band  in Interval III, as shown in Extended Data Fig. \ref{fig:SED_int3} (For more details of the synchrotron and SSC emission and its modeling, see Methods). 
Thus, our results indicate that GRB~180720B is the very first GRB event showing the apparent SSC components observed from the reverse shock in the early phase and the forward shock in the late phase.
In either case, the required magnetic parameter is low, $\epsilon_{B} \sim$  10$^{-4}$--10$^{-3}$ (See Methods and Extended Data Table \ref{table:parameters_rs_fs}). 
Although some previous works predicted a larger magnetization of the reverse shock compared with the forward shock due to injection of a magnetized ejecta from the fireball\cite{2013ApJ...772..101H}, 
our model requires the estimated magnetization of the forward and reverse shocks to be within the same 
order of magnitude, to be able to explain the strong SSC emission.
Thus, the high-energy $\gamma$-ray emission provides interesting constraints on the magnetization of the 
ejecta and the polarization measurement reveals the magnetic structure and its origin  in the shocked regions.

The scenario with the reverse shock emission is supported by the measurement of optical polarization.
At $t_{\rm obs} \sim T_0 + 80-300\;$s, when emission from the reverse shock dominates the flux, the PD changes gradually from $\sim5\%$ to $\lesssim1\%$ while the PA remains roughly constant at a mean value of $\sim70^\circ$. During $t_{\rm obs} \sim T_0 + 300-2000\;$s, when the lightcurve undergoes a transition from being reverse shock to forward shock dominated, the PD varies between $\sim2\%$ and $\sim8\%$ and the PA changes gradually and continuously. At late times ($t_{\rm obs} \gtrsim T_0 + 5000\;$s), when the forward shock dominates the total flux (by a factor of $\gtrsim$10 at $t_{\rm obs} \sim T_0 + 10^4\;$s, due to the steep temporal index of the reverse shock) as well as the polarized flux, the PD varies between $\sim0.5\%-2\%$ and the PA shows small fluctuations around its mean value of $\sim160^\circ$. This PA is different from that of the early reverse shock dominated emission by $\sim$90$^\circ$.

\begin{figure*}[ht!]   
\begin{center}
\hspace*{-15em} % move the table to left
\includegraphics[width=0.35\textwidth,viewport=0 0 610 1300]{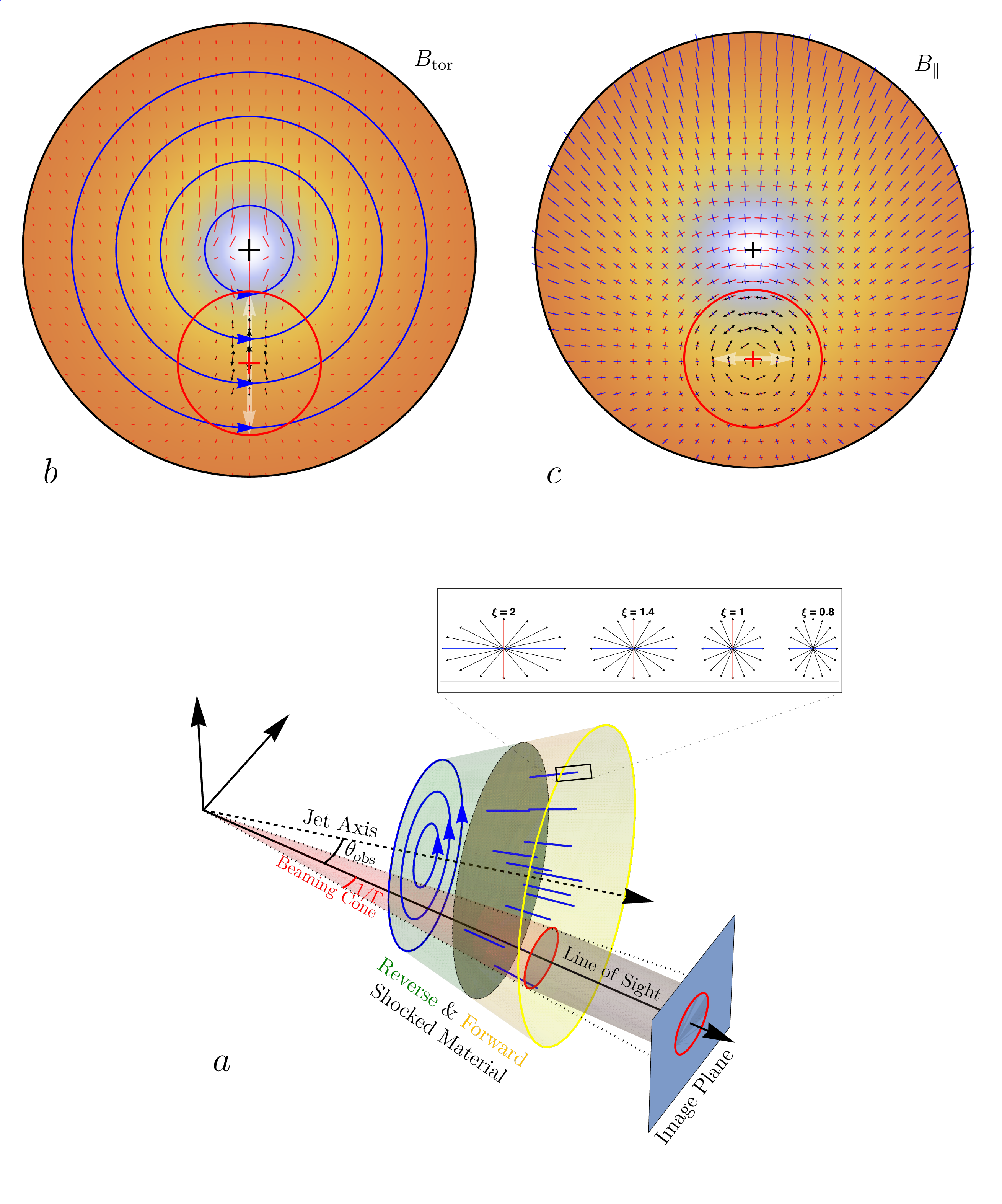}
\caption{\textbf{
Our polarization model.}
The early ($t_{\rm obs} \lesssim T_0 + 300\;$s) and late ($t_{\rm obs} \gtrsim T_0 + 5000\;$s) flux and polarization are dominated by emission from the reverse and forward shock regions, respectively (panel \textbf{a}; green and yellow shaded regions). In our model for the measured optical polarization the reverse shock region has a large-scale toroidal magnetic field, originating from the central source, centered on the jet's symmetry axis (circular blue lines with arrows in panels \textbf{a} and \textbf{b}), leading to a relatively large polarization degree, PD$\,\sim1\%-5\%$, with a polarization angle along the direction from the line of sight (red '+') to the jet axis (black '+'; panel \textbf{b}) at early times. The forward shock region has a shock-generated magnetic field that is somewhat smaller along the shock normal ($B_\parallel$) than perpendicular to it ($B_\perp$) just behind the shock, $\xi=B_\parallel/B_\perp<1$, but becomes larger along the shock-normal (i.e. radial direction; blue straight lines in panel \textbf{a} and in projection in panel \textbf{c}), $\xi>1$, in most of this region. This occurs since $\xi$ increases with the distance behind the forward shock because of the larger radial stretching of the shocked plasma\cite{Granot-Konigl-03,Gill-Granot-20} (inset above panel \textbf{b}). This results in a polarization angle perpendicular to the direction from the line of sight to the jet axis, with a relatively small PD$\,\sim0.5\%-2\%$ (panel \textbf{c}). In panels \textbf{b} and \textbf{c} the colormap signifies brighter emission closer to the jet axis,
the short red lines show the local polarization direction, the double-sided black arrows indicate the local polarized intensity, the red circle of angle $1/\Gamma_{\rm bulk}$ around the line of sight 
contains the region dominating the observed flux and polarization; the shaded white double-sided arrow represents the direction and relative strength of the net polarization from the angularly unresolved source. In addition to the magnetic fields described above, there is evidence in favor of randomly oriented small-size coherent magnetic field patches (possibly from turbulence or instabilities) that causes large variations in the polarization degree and direction at intermediate times ($t_{\rm obs} \sim T_0 + 300-2000\;$s) and reduce the otherwise larger degree of polarization at early times.
}
\label{fig:pol_model} 
\end{center}   
\end{figure*}

The early ($t_{\rm obs}\lesssim T_0 + 300\;$s) ejecta-dominated emission with a relatively high PD and 
roughly constant PA may originate from a combination of a large-scale transverse ordered magnetic field and a random field 
(e.g. from shock microphysical instabilities or turbulence), where the former dominates the polarized flux and the latter 
dominates the total flux.\cite{Granot-Konigl-03} The late ($t_{\rm obs} \gtrsim T_0 + 5000\;$s) emission is 
afterglow-dominated not only in terms of the total flux but also in its polarized flux. The fact that its PA is $\sim$90$^\circ$ 
different from the early-time PA may therefore be of great significance, as it relates, for the first time, between the 
magnetic field structures in the ejecta and in the shocked external medium. For example, for the commonly invoked large-scale 
toroidal magnetic field in the ejecta, symmetric around the jet symmetry axis, the early PA would be along the direction 
from our line-of-sight to the jet axis (see Fig.~\ref{fig:pol_model}). For an afterglow shock-produced magnetic field the polarization would be in the 
same direction if it were primarily random in the plane of the shock, as is usually assumed  based on theoretical considerations
of plasma instabilities. \cite{1999ApJ...526..697M,2009ApJ...693L.127K} 
However, it can be exactly $90^\circ$ different from this direction if it is more isotropic just after the shock and then becomes 
predominantly parallel to the shock normal due to larger stretching along this direction, where the latter dominates the total 
volume-averaged polarization by a small margin.\cite{Granot-Konigl-03,Gill-Granot-20} An observation like the one presented 
here is crucial for distinguishing between these two shock-generated field configurations (shock-normal or shock-plane dominated). 
At intermediate times ($t_{\rm obs} \sim T_0 + 300-2000\;$s) the above-described polarization from the two emission regions 
nearly cancel each other out, allowing an additional stochastic component 
%, e.g. from turbulence or instabilities, 
to dominate the polarized flux, leading to continuous short-timescale variation in PA and PD. This stochastic component may 
arise from turbulent magnetic fields that are coherent on hydrodynamic scales, and which can be produced, e.g., in shocks\cite{Gruzinov-Waxman-99}, 
by the Richtmyer-Meshkov instability due to density fluctuations,\cite{2007ApJ...671.1858S, 2011ApJ...734...77I} and the 
Rayleigh-Taylor instability at the contact discontinuity.\cite{2014ApJ...791L...1D}

The multi-wavelength observations with polarization measurements of GRB~180720B detected emissions from internal, external forward and reverse shocks, which are major candidates for $\gamma$-ray emission regions in GRBs (See Supplementary Methods for the internal shock). Our results strongly suggest that the early GeV emission is the first robust detection of SSC from the reverse shock where it is directly correlated to the optical emission with the polarization information, while the later TeV emission is SSC from the forward shock. 
An SSC origin of early GeV to TeV emission may help resolve the difficulties with a synchrotron origin, namely the violation of the maximum allowed synchrotron photon energy\cite{2014Sci...343...42A,2019Natur.575..455M, 2019Natur.575..459M}.
Furthermore,  our optical polarization measurements can help elucidate the origin and structure of GRB
magnetic fields, which are tightly coupled  to the particle acceleration mechanism.

\clearpage

%%%%%%%%%%%%%%%%%%%%%%%%%%%%%%%%%%%%%%%%%%%%%%%%%%%%%%%%%%%%%%%%%%%%%%%%%%%%%%%
%%%%%%% to count the word, uncomment the next line \end{document} and use the Overlef Menu-. Word Count
%\end{document}
%%%%%%%%%%%%%%%%%%%%%%%%%%%%%%%%%%%%%%%%%%%%%%%%%%%%%%%%%%%%%%%%%%%%%%%%%%%%%%%
\clearpage
\section*{Methods}
\subsection*{GeV and TeV $\gamma$-ray observations}
\Fermi-LAT\cite{2009ApJ...697.1071A} data  were processed with the Fermitools 1.2.23 and the event class of ``P8R3\_TRANSIENT020E\_V2'' was used to calculate $\gamma$-ray flux  with a power-law function. Here, the region of interest radius is 15$^\circ$ and the max zenith angle is 100$^\circ$.
\Fermi-LAT observation detects GeV $\gamma$-ray emission of this burst lasting for $\sim$1000 s. The GeV onset time is at $T_0$ + $\sim$100 s and the GeV flux declines with a temporal index of -1.91$\pm$0.31. After $T_0$ + 900 s, the GRB position was out of field of view of LAT with a source off-axis angle of 70$^\circ$.

The HESS flux area was derived from their observation paper\cite{2019Natur.575..464A} and the effect of the extra-galactic background light was corrected.

\subsection*{X-ray observations}
X-ray data of \Swift-XRT for lightcurves and spectra in this paper were processed by an automatic analysis procedure\cite{2009MNRAS.397.1177E}.
The BAT lightcurve declined with a temporal index of -1.79 $\pm$ 0.02 for $T_0$ + 20 s -- $T_0$ + 270 s and  after that a temporal index of -0.91 $\pm$ 0.17. For $T_0$ + 30 s -- $T_0$ + 270 s, the BAT lightcurve shows a short variability timescale with $\Delta t < t_{\rm obs}$ and a strong spectral evolution, which indicates that the BAT lightcurve likely originates from an internal shock\cite{2006ApJ...641.1010F}.

For XRT data, the initial bright X-ray flare observed at $T_0$ + 100 s  has rise and decay temporal indices of +4.85 $\pm$ 0.44 and -8.18 $\pm$ 0.3, respectively, and a hard photon index ($\Gamma_{\rm ph}=-1.5\pm0.1$, see the 
third panel of Fig. \ref{fig:LC}).  Such a short variability timescale indicates that the X-ray flare does not originate from the afterglow emission site -- the external shock.  Most likely the X-ray flare originates from a different site, e.g., an internal shock.
The underlying power-law component has temporal indices of -1.22$\pm$0.02 ($T_0$ + 80 s to $T_0$ + 380 s), -0.75$\pm$0.01 ($T_0$ + 380 s to $T_0$ + 2780 s),  -1.28$\pm$0.02 ($T_0$ + 2780 s to $T_0$ + 1.5 $\times$ 10$^5$ s), and -1.58$\pm$0.03 ($T_0$ + 1.5 $\times$ 10$^5$ s to $T_0$ + 4.0 $\times$ 10$^4$ s). For $T_0$ + 380 s to $T_0$ + 2780 s, the temporal index is much shallower than a typical one (e.g., -1.2), called shallow decay.  One of the possible origins is energy injection from a central engine\cite{2006ApJ...642..389N,2006ApJ...642..354Z}, 
which makes the forward shock emission decay more slowly. Such energy injection may be either due to a long-lived central engine producing a relativistic wind for a long time, or a short-lived central engine that produces an outflow with a wide range of Lorentz factors where slower and more energetic matter resides behind faster moving matter, eventually catching up with it and energizing it.\cite{1998ApJ...496L...1R,2006ApJ...642..354Z,2009MNRAS.395.1941M}

In the time interval from $T_0$ + 2780 s to $T_0$ + 1.5 $\times$ 10$^5$ s, the observed temporal index of -1.28$\pm$0.02 can be interpreted as an normal decay phase in the standard forward-shock afterglow theory\cite{1998ApJ...497L..17S}. There is a temporal break at $T_0$ + 1.5 $\times$ 10$^5$ s ($\pm$ 0.2 $\times$ 10$^5$ s) and the difference of the temporal indices is $\sim$0.3. One of the possible break candidates is the cooling break in which the synchrotron cooling frequency passes through the observed frequency.  Such a scenario predicts that the photon index becomes softer by 0.5\cite{1998ApJ...497L..17S}. However, after $T_0$ + 1.5 $\times$ 10$^5$ s the observed photon index ($\Gamma_{\rm ph} \sim$ -1.7) is almost constant as a function of time (see the middle panel of Fig. \ref{fig:LC}). Thus, the temporal break at $T_0$ + 1.5 $\times$ 10$^5$ s is not caused by the cooling break, but another jet effect (e.g., jet break in which the relativistic jet is decelerated and the side expansion becomes significant).

\subsection*{Optical photometric observations}
The optical flux data points obtained by the Kanata and other ground telescopes reported in the GCN circular were used\cite{2018GCN.22977....1S, 2018GCN.22979....1R,2018GCN.22983....1I, 2018GCN.22985....1K,2018GCN.22988....1C,2018GCN.23004....1H, 2018GCN.23020....1S, 2018GCN.23033....1Z}.
The Kanata observations were performed with imaging polarimetry mode on the first day (see Subsection ``Optical polarimetric observation'') and with imaging mode on the second and the third days. All the Kanata photometric data have been calibrated by the relative photometry to nearby stars using the APASS catalog\cite{henden2016vizier}.
Phenomenologically the optical light curve is represented as a simple power-law component for reverse shock plus a power-law component with two temporal breaks for forward shock.
For the former power-law component, the best-fit temporal index is -1.95$\pm$0.02. For the latter power-law component consists of -0.31$\pm$0.01 (before $T_0$ + 8300 s), -1.10$\pm$0.02 ($T_0$ + 8300 s -- 1.3 $\times$ 10$^5$ s) and -1.94$\pm$0.08 (after $T_0$ + 1.3 $\times$ 10$^5$ s). 
For the time interval before $T_0$ + 8300 s, the shallow temporal index may indicate the same origin of the X-ray emission in shallow decay phase described in the previous section, but the sparse data points limit further detailed discussion on this.  For $T_0$ + 8300 s -- 1.3 $\times$ 10$^5$ s, the observed temporal index of  -1.10$\pm$0.02 is almost similar to the observed X-ray temporal index (-1.28$\pm$0.02) in the same time interval. 
For the time interval before $T_0$ + 1.3 $\times$ 10$^5$ s, the temporal index becomes steeper than the previous time section. The optical temporal break at (1.3 $\pm$ 0.2) $\times$ 10$^5$ s occurred simultaneously with the X-ray temporal break at $T_0$ + (1.5 $\pm$ 0.2) $\times$ 10$^5$ s, which suggests that the achromatic temporal break at $\sim$1.3 $\times$ 10$^5$ s originates from the relativistic effect of the side expansion, called jet break.

\subsection*{Spectral energy distributions in Intervals I and II}

For the SEDs from $T_0$ + 80 -- 130 s  (Interval I) and $T_0$ + 160 -- 300 s (Interval II)  where the bright and temporally steep optical and GeV emission was observed, the joint-fit results of the \Swift-XRT and \Swift-BAT show that the best-fit function favors  the broken power-law function or the Band function significantly compared with the simple power-law function over the time interval I and II, as shown in Supplementary Table 
\ref{table:spectrum_XRT_BAT}.
Note that there is no significant difference between the broken power-law function and the Band function. In the interval II, the broken power-law function is marginally favored and we adopt the broken power-law function to represent the X-ray component in the main texts. We also find that the choice of the two function does not significantly affect the extrapolation to GeV energies.

For the optical data analysis, we adopt the extinction of the host galaxy to be $A_v$ = 0.5 mag as a typical case \cite{2012A&A...537A..15S}. 
The obtained SEDs from the optical band to the GeV band is shown in  Fig. \ref{fig:SED_multi}. In the time interval I ($T_0$ + 80 -- 130 s) where the X-ray flare occurred, the optical component is distinctly higher than the extrapolation from the X-ray component being well fitted by the broken power-law function, while 
the excess of the GeV component over the X-ray extrapolation is not significant, which is due to an additional contribution from the X-ray flare likely originating from an internal shock. 
In the time interval II ($T_0$ + 160 -- 300 s), after the occurrence of the X-ray flare, the SED shows that the optical and GeV fluxes are significantly larger than the extrapolations from the best-fit broken power-law function in the X-ray band ($>$5 $\sigma$ and $\sim4 \sigma$ confidence levels in the optical and GeV bands, respectively. For more details of the GeV excess, see the subsection of Methods ``Statistical significance of the GeV excess''). This indicates that both the optical and GeV components have different origins from the X-ray component. Furthermore, the initial steep temporal index of $\alpha_{\rm opt} \sim \alpha_{\rm GeV} \sim$ -1.9 suggests that the optical and the GeV components have the same origin and the steep temporal index can be interpreted as having a reverse-shock origin\cite{2000ApJ...545..807K}.

\subsection*{Theoretical modeling of the forward shock at the late afterglow phase (``analytical model'')}

The emission from the forward shock, which is a different emission component from the reverse-shock emission, is not the main topic in this paper. For reference, however, we show the ``analytical'' modeling for the forward shock emission. The X-ray lightcurve, which is dominated by the forward shock emission, shows complex behavior, so that a modeling with a simple model is unfortunately hard to reconcile the observed data.

To constrain the shock evolution, we first focus on the time interval III ($T_0$ + $4.5 \times 10^4$ s), in which the VHE $\gamma$-ray data was taken with the HESS.
In this phase, the emission is dominated by the forward shock emission.
By extracting the SED in time interval  IV (Supplementary Fig. \ref{fig:SED_int4}), 
we find that the X-ray and optical components are on the 
same power-law segment with a photon index of $\Gamma_{\rm ph}$ = -1.8. 
To explain the observed temporal and spectral indices ($\alpha_{\rm X} \sim$ -1.3 and $\Gamma_{\rm ph} \sim$ -1.8), 
a  circumburst medium with a wind profile ($n (R) \propto R^{-2}$) is in tension with the fast or slow cooling synchrotron scenario.
Assuming a constant interstellar medium 
($n_{\rm ISM} (R) \propto R^{0}$), the optical and X-ray temporal decay indices of $\alpha \sim$ -1.3 and the photon spectral index of 
$\Gamma_{\rm ph} \sim -1.8$ suggest that a power-law index of the electron injection spectrum is $p \sim 2.7$ ($\alpha = 3(1-p)/4$ and 
$\Gamma_{\rm ph} = -(1+p)/2$ for $\nu_{\rm m} < \nu_{\rm obs} < \nu_{\rm c}$, where $\nu_{\rm m}$ is the typical synchrotron frequency and $\nu_{\rm c}$ is the synchrotron cooling frequency.).
The radio flux point suggests that the spectral break at $\nu_{\rm m}$ should exist between
the radio and optical bands. The number density of the interstellar medium should be as small 
as $n_{\rm ISM} \sim 10^{-3}\,{\rm cm}^{-3}$, because the synchrotron cooling frequency 
($\nu_{\rm c} \propto n_{\rm ISM}^{-1} \epsilon_{B}^{-3/2}$) should be above the XRT band  ($\sim$ a few keV).
 In addition, $\epsilon_{B}$ is roughly constrained due to $\nu_{\rm m}$ ($\propto n_{\rm ISM}^{0} \epsilon_{B}^{1/2}$) being located between the radio and the optical bands (and also for compatibility with the radio flux) and to obtain the correct ratio of the SSC to synchrotron emission ($Y$). Thus, this GRB needs to have an atypically low $n_{\rm ISM}$ of $\sim 10^{-3}$ cm$^{-3}$. Such a low $n_{\rm ISM}$ value is not so unlikely and even lower values of $\sim$ $10^{-4}-10^{-3}$ cm$^{-3}$ have been reported for some GRBs with multi-wavelength observations.\cite{2015ApJ...814....1L, 2018ApJ...862...94L}

The SSC characteristic frequencies and flux can be derived from the synchrotron ones multiplied by $\gamma_{\rm c}^2$ and the Compton $Y$ parameter, respectively. When the Klein Nishina effect is neglected\cite{2001ApJ...548..787S}, $Y$ = ($\epsilon_{\rm e}/\epsilon_{B}$)$^{1/2}$ ($\gamma_{\rm c}/\gamma_{\rm m}$)$^{(2-p)/2}$, where $\gamma_{\rm m}$ and $\gamma_{\rm c}$ are the Lorentz factors of minimal energy electrons and those that are cooling at the dynamical time.
Following the analytical formulation \cite{2009ApJ...703..675N,2022MNRAS.tmp..517Y}, we model the synchrotron and inverse Compton components 
at this time interval.
Given a snapshot of the spectrum at a certain observation time $t_{\rm obs}$, the fitting parameters are five:
the electron spectral index $p$, the electron minimum Lorentz factor $\gamma_{\rm m}$, the ratio of the energy fractions of non-thermal electrons to the magnetic field $\epsilon_{\rm e}/\epsilon_{B}$, the magnetic field $B$, and the bulk Lorentz factor $\Gamma_{\rm bulk}$.
As summarised in Extended Data Table \ref{table:parameters_rs_fs} ($t_{\rm obs}=4.5\times10^4$ s), we adopt $\epsilon_{\rm e}/\epsilon_{B}=1330$.
Taking into account the Klein-Nishina effect, the electron Lorentz factor at the cooling break $\gamma_{\rm c}$ ($>\gamma_{\rm m}$, slow cooling case)
and the Compton $Y$ parameter at $\gamma_{\rm c}$ ($Y_{\rm c}$) are obtained from
\begin{eqnarray}
(1+Y_{\rm c})\gamma_{\rm c}&=&\frac{6 \pi m_{\rm e}c (1+z)}{\sigma_{\rm T} B^2 \Gamma_{\rm bulk} t_{\rm obs}}, \\
Y_{\rm c} (1+Y_{\rm c})&=&\frac{\epsilon_{\rm e}}{\epsilon_{B}}
\left( \frac{\gamma_{\rm c}}{\gamma_{\rm m}} \right)^{2-p}
\left( \frac{\hat{\gamma}_{\rm c}}{\gamma_{\rm c}} \right)^{(3-p)/2},
\end{eqnarray}
where the Lorentz factor $\hat{\gamma}_{\rm c} \equiv 0.2 \Gamma_{\rm bulk} m_{\rm e}c^2/((1+z)h\nu_{\rm c})$, above which the Klein-Nishina effect prevents electrons from upscattering synchrotron photons at the spectral peak $\nu_{\rm c} \equiv \Gamma_{\rm bulk} \gamma_{\rm c}^2 e B/(2 \pi m_{\rm e} c(1+z))$.
Our model parameters yield $\gamma_{\rm c}=1.7\times 10^6$,
$Y_{\rm c}=2.1$, and $\hat{\gamma}_{\rm c}=230$.
Since $\hat{\gamma}_{\rm c}<\gamma_{\rm m}<\gamma_{\rm c}<\hat{\gamma}_{\rm m}\equiv 0.2 \Gamma_{\rm bulk} m_{\rm e}c^2/((1+z)h\nu_{\rm m})=2 \times 10^7$, electrons between $\gamma_{\rm m}$ and $\hat{\gamma}_{\rm m}$ 
can scatter photons with frequency of $\nu_{\rm m}<\nu<\nu_{\rm c}$. The electron spectrum below $\gamma_{\rm c}$ is $N(\gamma)\propto \gamma^{-p}$
so that the synchrotron spectrum between $\nu_{\rm m} \equiv \Gamma_{\rm bulk} \gamma_{\rm m}^2 e B/(2 \pi m_{\rm e} c(1+z))$ and $\nu_{\rm c}$ 
is $F_\nu \propto \nu^{-(p-1)/2}$. Let us consider that electrons with $\gamma>\gamma_{\rm c}$ cool via SSC emission mainly scattering target photons of
$\nu_{\rm t} \simeq 0.2 \Gamma_{\rm bulk} m_{\rm e}c^2/((1+z)\gamma)$,
though the SSC and synchrotron cooling rates are comparable at $\gamma>\gamma_{\rm c}$.
In this case, the energy loss rate
can be approximated as $\dot{\gamma} 
\propto \gamma^2 \nu_{\rm t} F_{\nu_{\rm t}}
\propto \gamma^{(p+1)/2}$.
This leads to the electron spectrum above $\gamma_{\rm c}$ as $N(\gamma) \propto \gamma^{-p+1}/\dot{\gamma} \propto
\gamma^{-(3p-1)/2}$, which yields the synchrotron spectrum above $\nu_{\rm c}$
as $F_\nu \propto \nu^{-3(p-1)/4}$.
The Compton Y-parameter decreases with energy as $Y \propto \nu_{\rm t} F_{\nu_{\rm t}} \propto \gamma^{(p-3)/2}$ and $\propto \gamma^{-4/3}$ for $\gamma<\hat{\gamma}_{\rm m}$ and $\gamma>\hat{\gamma}_{\rm m}$, respectively. The synchrotron spectrum would show a structure at $\nu=\nu_0$ that is the typical synchrotron frequency emitted by electrons of $\gamma=\gamma_0=2.6 \times 10^7$ at which $Y=1$. We omit detailed discussion about such structures at $\nu=\hat{\nu}_{\rm m}$ or $\nu_0$.

The peak of the SSC spectrum is at $\nu=\nu_{\rm IC}=2 \gamma_{\rm c}
\hat{\gamma}_{\rm c} \nu_{\rm c}$,
which corresponds to $0.4 \gamma_{\rm c} m_{\rm e}c^2$
in the shocked-fluid rest frame.
The typical frequency of scattered photons is $\nu \sim \gamma^2 \nu_{\rm t} \propto \gamma$. Then, the SSC spectrum, $F_\nu \propto \gamma N(\gamma) \gamma^2 \nu_{\rm t} F_{\nu_{\rm t}}/\nu$ with $\nu_{\rm t} \propto \nu^{-1}$, provides
$F_\nu \propto \nu^{-(p-1)/2}$
and $\propto \nu^{-p+1}$ for $2 \gamma_{\rm m}^2 \nu_{\rm m}<\nu<\nu_{\rm IC}$ and $\nu > \nu_{\rm IC}$, respectively. Our parameter set gives us $\nu_{\rm m}=2.5 \times 10^{13}$ Hz, $\nu_{\rm c}=2.1 \times 10^{18}$ Hz, $2 \gamma_{\rm m}^2 \nu_{\rm m}=1.7 \times 10^{21}$ Hz, and $\nu_{\rm IC}=1.7 \times 10^{27}$ Hz.

The electron energy density obtained from the total energy $(p-1)N_{\rm e} \gamma_{\rm m} m_{\rm e}c^2/(p-2)$, and the volume of the shocked ISM, $\pi R^3/(3 \Gamma_{\rm bulk})$, where the radius $R=4 c t_{\rm obs}\Gamma_{\rm bulk}^2/(1+z)$, is equivalent to $(\epsilon_{\rm e}/\epsilon_{B}) B^2/(8 \pi)$, from which the total electron number $N_{\rm e}$ is written with our five model parameters.
The synchrotron flux at $\nu=\nu_{\rm m}$ in this slow cooling case is written not apparently depending on both the total energy and the ISM density as
\begin{eqnarray}
F_{\nu_{\rm m}}=\Gamma_{\rm bulk} \frac{1+z}{4\pi d_{\rm L}^2}  \frac{\sigma_{\rm T} m_{\rm e}c^2 B}{3e}N_{\rm e}=\frac{p-2}{p-1}\frac{\epsilon_{\rm e}}{\epsilon_{B}}\frac{2 \sigma_{\rm T} c^3 t_{\rm obs}^3 B^3 \Gamma_{\rm bulk}^6}{9 \pi (1+z)^2 d_{\rm L}^2 e \gamma_{\rm m}}.
\label{eq:Fnum}
\end{eqnarray}
Our choice of the parameters gives us $\nu_{\rm m} F_{\nu_{\rm m}}=3.0 \times 10^{-12}~\mbox{erg}~\mbox{cm}^{-2}~\mbox{s}^{-1}$. Normalized with this value, we plot the model spectrum in Extended Data Fig. \ref{fig:SED_int3}.
The peak ratio of the SSC component to the synchrotron component in $\nu F_\nu$-plot is given as $Y_{\rm c}$.

From the parameter set, we obtain $\epsilon_{\rm e}/f_{\rm e}=(p-1) \gamma_{\rm m} m_{\rm e}/((p-2)\Gamma_{\rm bulk} m_{\rm p})=0.24$, where $f_{\rm e}$ is the number fraction of non-thermal electrons, and $\epsilon_{B} n_{\rm ISM}=B_{\rm f}^2/(32\pi m_{\rm p} c^2 \Gamma_{\rm bulk}^2)=1.1 \times 10^{-6}~\mbox{cm}^{-3}$. The implied value $f_{\rm e} n_{\rm ISM}=6.0\times 10^{-3}~\mbox{cm}^{-3}$ suggests a relatively low ISM density. 
Assuming $\epsilon_{\rm e}$ = 0.2, we obtain $\epsilon_{B}$ = 1.5 $\times$ 10$^{-4}$, $f_{\rm e}=0.82$, and $n_{\rm ISM}$ = 7.3 $\times$ 10$^{-3}$ cm$^{-3}$. Those parameters can well reproduce the observed SED in time interval III as shown in Extended Data Fig. \ref{fig:SED_int3}.
The adopted  parameters of the synchrotron shock model (see ``analytical'' in  Extended Data Table \ref{table:parameters_rs_fs}) are mostly consistent with previous works of GRB~180720B%\cite{2019ApJ...884..117W, 2019ApJ...885...29F}. 
\cite{2019ApJ...884..117W}. 
 The modeled lightcurves of the forward shock at different frequencies are shown in Extended Data Fig. \ref{fig:LC-model}.

\subsection*{Theoretical modeling of the early afterglow with the reverse shocked component (``model 1/2'')}

In the time interval II, the X-ray lightcurve behaves like that in the typical shallow decay phase, while the optical and the GeV $\gamma$-ray components
steeply decay. First, we model the X-ray component with the forward shock emission as shown in Fig. \ref{fig:SED_multi} ($t_{\rm obs}=200$ s).
The model is partially constrained by the spectral model for the time interval III. With the one-zone approximation \cite{2017ApJ...844...92F}, the flux constrained by equation (\ref{eq:Fnum}) gives a constant ratio
\begin{eqnarray}
\hat{E} \equiv \frac{E_0}{n_{\rm ISM}}
=\frac{4^4 \pi m_{\rm p} c^5 \Gamma_{\rm bulk}^8}{3(1+z)^3} t_{\rm obs}^3=2.3\times 10^{56}~\mbox{erg}~\mbox{cm}^3,
\label{eq:Ehat}
\end{eqnarray}
where $E_0$ is the total kinetic energy.
If we assume a constant $n_{\rm ISM}$, $\Gamma_{\rm bulk} \simeq 240$ at $t_{\rm obs}=200$ s.
However, if we maintain the microscopic parameters adopted for the time interval III ($t_{\rm obs}=4.5 \times 10^4$ s) even at $t_{\rm obs}=200$ s, the X-ray flux from the forward shock is not reproduced. In this period, the X-ray lightcurve is in the shallow decay phase, but the energy injection model, which implies internal collisions from behind, is not preferable for the steeply decaying reverse shock emission. The internal collisions should inject energy to the reverse-shocked region too. In addition, the X-ray spectrum shows a break at $\sim 3$ keV. If the microscopic parameters are constant, $\nu_{\rm m}$ is still lower than 3 keV, and we obtain $\gamma_{\rm m} \simeq \hat{\gamma}_{\rm m}$ heavily suppressing the IC cooling of electrons above $\gamma_{\rm m}$.

We need temporal evolution of the microscopic parameters of the forward shock to make a cooling break at $\sim 3$ keV. One example of such models is shown in Fig. \ref{fig:SED_multi}, where the two components, the forward and reverse shock components, are shown. Hereafter, we denote the model parameters for the forward (reverse) shock with subscript ``f'' (``r'').
For the forward shock component, we adopt $\Gamma_{\rm bulk,f}=240$, $p=2.2$, $\epsilon_{\rm e,f}/\epsilon_{B,{\rm f}}=1.33$, $\gamma_{\rm m,f}=6700$, and $B_{\rm f}=0.622$ G. The same calculations as those for the late afterglow yield $\hat{\gamma}_{\rm c,f}=4700<\gamma_{\rm c,f}=5.5 \times 10^4<\hat{\gamma}_{\rm m,f}=3.2 \times 10^5$, $\nu_{\rm m,f}=1.1 \times 10^{16}$ Hz, $\nu_{\rm c,f}=7.6 \times 10^{17}$ Hz, $2 \gamma_{\rm c,f} \hat{\gamma}_{\rm c,f} \nu_{\rm c,f}=3.9 \times 10^{26}$ Hz, $Y_{\rm c,f}=0.26$, and $\nu_{\rm m,f} F_{\nu_{\rm m,f}}=8.9 \times 10^{-10}~\mbox{erg}~\mbox{cm}^{-2}~\mbox{s}^{-1}$.
The parameter set implies $\epsilon_{\rm e,f}/f_{\rm e,f}=0.09$, and $\epsilon_{B,{\rm f}} n_{\rm ISM}=4.4 \times 10^{-5}~\mbox{cm}^{-3}$. The implied value $f_{\rm e,f} n_{\rm ISM}=6.5\times 10^{-4}~\mbox{cm}^{-3}$ with $n_{\rm ISM}$ = 7.3 $\times$ 10$^{-3}$ cm$^{-3}$ suggests a lower $f_{\rm e}$ ($\sim 0.09$) compared to the value in the interval II and $\epsilon_{\rm e,f} \sim \epsilon_{B,{\rm f}} \sim$ 10$^{-2}$.

Next, we move on to the emission from the shocked ejecta. After the reverse shock crosses the shell, the Blanford--McKee solution\cite{1976PhFl...19.1130B} is not applicable for the shocked ejecta. A general power-law scaling law of $\Gamma_{\rm bulk} \propto R^{-g}$ \cite{2000ApJ...542..819K} is useful to characterize the evolution of physical parameters of a reverse shock, where $g$ is the constant ranging 3/2 $<$ $g$ $<$ 7/2 for the ISM case.
In this case, the temporal index at the observed band for $\nu_{\rm m,r} < \nu_{\rm obs} < \nu_{\rm c,r}$ is given as $\alpha_{\rm RS}^{\rm syn} = -[(15g+24)p + 7g]/(28g + 14)$.  Adopting $p= 2.35$ and $g$ = 5/2, we obtain $\alpha_{\rm RS}^{\rm syn} \sim -1.9$, 
 which is consistent with the observed temporal index of the optical emission.
Note that the observed temporal index does not depend on the $g$ index strongly.
 The temporal index of the SSC emission of the reverse shock is expected to be  
 %$-(p+1)/2 \sim -1.7$ \cite{2016ApJ...818..190F} 
 $\alpha_{\rm RS}^{\rm SSC} = [-3g+26 - p(19g + 36)]/[14(2g + 1)] \sim -2.1$ \cite{2020ApJ...905..112F}
 and this value is also consistent with the observed temporal index of the GeV $\gamma$-ray emission. 

Here, we also check whether the temporal index of the LAT emission is compatible with the other model, e.g., the SSC component from the forward-shock emission.  
For $\nu_{\rm m, f}^{\rm SSC} < \nu < \nu_{\rm c, f}^{\rm SSC}$ (or $\nu_{\rm max, f}^{\rm SSC}$) in the slow cooling regime, the theoretical SSC temporal indices of the forward shock are $(11-9p)/8 \sim -1.27$ and $-p = -2.35$ in the ISM and wind profiles, respectively, for $p$ = 2.35.\cite{2000ApJ...543...66P,2019ApJ...883..162F} 
Since the ISM profile is favored (see ``Theoretical modeling of the forward shock at the late afterglow phase'' in Methods), the temporal index of the theoretical SSC lightcurve is inconsistent with the observed temporal index of the LAT lightcurve ($-1.9\pm$0.3). Thus, the scenario that the LAT lightcurve arises from the forward shock may be rejected. Note that, more conservatively, if we do not determine or constrain the electron index ($p$) or the external medium density profile we cannot strongly constrain whether the LAT emission originates from the forward or reverse shock. 
Thus, only the temporal information in the GeV band cannot put the strong constraint on the emission site and the temporal information from
the multi-wavelength observations in the early to late phase is crucial.

The optical and $\gamma$-ray lightcurves suggest that the onset of the reverse shock emissions is earlier than $t_{\rm obs}=100$ s,
at which $\Gamma_{\rm bulk} \simeq 310$ from equation (\ref{eq:Ehat}). This is consistent with the deceleration time \cite{2017ApJ...844...92F}
\begin{eqnarray}
t_{\rm dec}=\frac{1+z}{2} \left( \frac{3 \hat{E}}{32 \pi m_{\rm p} c^5 \Gamma_0^8} \right)^{1/3}\simeq 100 \left( \frac{\hat{E}}{2.3 \times 10^{56}~\mbox{erg}~\mbox{cm}^3} \right)^{1/3} \left( \frac{\Gamma_0}{310} \right)^{-8/3}~\mbox{s},
\label{eq:dec_time}
\end{eqnarray}
where $\Gamma_0$ is the initial Lorentz factor of the ejecta.
If the shocked ejecta starts deceleration ($\Gamma_{\rm bulk,r} \propto R^{-5/2} \propto t_{\rm obs}^{-5/12}$) at $t_{\rm obs}=100$ s, we obtain $\Gamma_{\rm bulk,r}=233$ at $t_{\rm obs}=200$ s.
The model parameters constrain the volume of the emission region. The implied width of the shocked region should be comparable to or larger than the typical shell width, $R/(12 \Gamma_{\rm bulk,r})$. However, a model with $\Gamma_{\rm bulk,r} \sim 200$ requires an extremely thin shell. We assume that the optical and $\gamma$-ray components are 
emitted from a slower ejecta, which may correspond to a fraction of the shocked ejecta decelerated by rarefaction wave. 
Thus, we adopt $\Gamma_{\rm bulk,r} = 90$ at $t_{\rm obs}= T_0 + 200$ s. 

The observed GeV spectrum has a photon index $\Gamma_{\rm ph}=-2.2\pm0.2$ which is softer than the theoretical SSC component that has $\Gamma_{\rm ph}=-(1+$p$)/2 = -1.7$, where $p\sim 2.35$ as discussed above. This suggests that the SSC component may have a spectral cutoff in the GeV band. Thus, 
we introduce an additional parameter, the maximum electron Lorentz factor $\gamma_{\rm max,r}$, which may be significantly lower than the values in the forward shock region. In Fig. \ref{fig:SED_multi}, we show two models for the reverse shock emission, which yield the same synchrotron spectrum. The model 1 has a similar $\epsilon_{\rm e}/\epsilon_{B}$ to the forward shock region in the interval III, while we adopt a more extreme value of $\epsilon_{\rm e}/\epsilon_{B}$ in the model 2 to maximize the emission volume.  The common parameters are $p=2.35$, and the maximum Lorentz factor $\gamma_{\rm max,r}=10 \gamma_{\rm m,r}$. 
Here, the maximum electron energy in the mildly relativistic reverse shock may be different from that in 
the ultra relativistic forward shock in GRBs. Recent PIC simulations\cite{2023arXiv230512008R} show very soft spectra of electrons accelerated 
by mildly relativistic shocks and the spectral softness  depends on the magnetic structure strongly.
Thus, the efficiency of the injection into the acceleration process in mildly relativistic shocks may be low. Alternatively, the turbulence acceleration rather than the shock acceleration may be the dominant process in the reverse shock region. Several simulations\cite{2013ApJ...775...87D,2020Galax...8...33V} show that the contact discontinuity between the forward and reverse shocks is likely unstable. The turbulence behind the forward shock may destroy the sharp reverse shock structure.

The other parameters are $\epsilon_{\rm e,r}/\epsilon_{B,{\rm r}}=1000$ (3300), $\gamma_{\rm m,r}=380$ (565), $B_{\rm r}=1.1$ G (0.52 G) for the model 1 (model 2). The cooling Lorentz factor due to synchrotron radiation is much larger than $\gamma_{\rm max,r}$ as $\gamma_{\rm c,r}=5.5 \times 10^4$ and $2.7 \times 10^5$ for the models with $\gamma_{\rm m,r}=380$ and 565, respectively. In these cases, the Klein--Nishina effect is negligible. As the synchrotron cooling time is proportional to $\gamma^{-1}$, we can approximate the Compton $Y$ parameter as $Y\simeq \sqrt{\epsilon_{\rm e,r}/\epsilon_{B,{\rm r}}} (\gamma_{\rm max,r}/\gamma_{\rm m,r})^{(2-p)/2} \sqrt{\gamma_{\rm max,r}/\gamma_{\rm c,r}}$. We have adjusted the parameter $\epsilon_{\rm e,r}/\epsilon_{B,{\rm r}}$ to the common value $Y=5.6$ for the two models. The parameter set yields $\nu_{\rm m,r}=2.5\times10^{13}$ Hz, $\nu_{\rm max,r}=2.5\times10^{15}$ Hz, $2 \gamma_{\rm max,r}^2 \nu_{\rm max,r}=7.2 \times 10^{22}$ Hz ($1.6 \times 10^{23}$ Hz) for the ``model 1 (model 2)'', where $\nu_{\rm max}$ is the maximum synchrotron frequency. The spectrum above $\nu_{\rm m}$ is written as $F_{\nu} \propto \nu^{-(p-1)/2} \exp{(-\sqrt{\nu/\nu_{\rm max}})}$. 
Note that taking into account the gradual cutoff shape above $\nu_{\rm max,r}$ of the SSC emission, the modeled SSC emission from the reverse shock reaches a few GeV as shown in Fig. \ref{fig:SED_multi}.

The number of non-thermal electrons $N_{\rm e,r}=f_{\rm e,r} E_{\rm ej}/(\Gamma_0 m_{\rm p} c^2)$,
where $E_{\rm ej}$ is the initial total energy of the ejecta
before transferring its energy to the forward shock, is constant during the emission. The flux is normalized as
\begin{eqnarray}
\nu_{\rm m,r} F_{\nu_{\rm m,r}}=\frac{1}{4 \pi d_{\rm L}^2}\frac{N_{\rm e,r} \sigma_{\rm T}cB_{\rm r}^2\gamma_{\rm m,r}^2 \Gamma_{\rm bulk,r}^2}{6 \pi}=7.2 \times 10^{-10} ~\mbox{erg}~\mbox{cm}^{-2}~\mbox{s}^{-1}.
\end{eqnarray}
Then, we obtain $E_{\rm ej}=4.0 (8.7) \times 10^{53}$ ($\Gamma_0/310$) $ f_{\rm e,r}^{-1}$ erg for the model 1 (model 2).
From $V_{\rm r}(\epsilon_{\rm e,r}/\epsilon_{B,{\rm r}}) B_{\rm r}^2/(8\pi)=(p-1) N_{\rm e,r} \gamma_{\rm m,r} m_{\rm e} c^2/(p-2)$, the volume of the emission region is estimated as $V_{\rm r}=2.0 \times 10^{49}~\mbox{cm}^3$ ($9.4 \times 10^{49}~\mbox{cm}^3$) for $\gamma_{\rm m,r}=380$ (565).
Since the emission radius is estimated as $R=4 \Gamma_{\rm bulk,r}^2 ct_{\rm obs}/(1+z)=1.2 \times 10^{17}$ cm, the shell width in model 1 (model 2) is estimated as $1.1 \times 10^{14}$ cm ($5.4 \times 10^{14}$ cm), which is close to (larger than) the typical width $R/(12 \Gamma_{\rm bulk}) \simeq 1.1 \times 10^{14}$ cm.
The adopted physical parameters used for reverse shock afterglow modeling are summarized in Extended Data Table \ref{table:parameters_rs_fs} (see ``model 1/2'').
The modeled lightcurves of the reverse shock at different frequencies are shown in Extended Data Fig. \ref{fig:LC-model}, where the models 1 and 2 are almost identical on the plot.

For the maximum SSC photon energy, 
if the inverse Compton scattering process occurs in the Thomson regime, the SSC spectrum extends up to $\gamma_{\rm c,r}^2\nu_{\rm c,r}$. 
A typical synchrotron cooling energy of the seed photon is $\nu_{\rm c,r} \sim 1\,\, {\rm keV} $ in the observer frame\cite{2000ApJ...545..807K}. Correspondingly, the seed photon energy in the rest frame of the electron is $\nu_{\rm seed}^\prime \sim (1+z)\gamma_{\rm c,r} \nu_{\rm c,r}/\Gamma_{\rm bulk,r} \sim$ 1 MeV ($\sim$ $m_{\rm e} c^2$), where $ \nu_{\rm c,r} \sim$ 1 keV,  $\gamma_{\rm c,r} \sim 10^5$ and $\Gamma_{\rm bulk,r} \sim$ 90 are derived from the above model.   This GRB has a 5-GeV photon event observed at $T_{\rm 0}$ + 142 s which was also reported in an earlier work\cite{2019ApJ...885...29F}. Although this event is slightly out of the time interval II, the SSC spectrum may be extended to above a few GeV energies, in which case model 2 with the higher-energy emission may be favored over model 1.  The high-energy cutoff energy of the SSC emission from the reverse shock $\nu_{\rm cutoff, r}^{\rm SSC}$ is determined by $\gamma_{\rm max, r}$. Thus, if we slightly modify $\gamma_{\rm max, r}$, the cutoff energy  can be extended to higher energies ($\nu_{\rm cutoff, r}^{\rm SSC} \propto \gamma_{\rm max,r}^2$).
The Klein-Nishina effect can strongly affect SSC photons on the high-energy side and the highest photon energy of the SSC component could be limited to $\Gamma_{\rm bulk,r}(1+z)^{-1} \gamma_{\rm c,r} m_{\rm e} c^2 \sim 10 (1+z)^{-1}\,$TeV, which is  larger than the LAT energy range.  Note that the seed photon energy for the second-order IC emission in the rest frame of the electron is $\nu_{\rm seed, SSC}^\prime \sim (1+z)\gamma_{\rm m,r} \nu_{\rm SSC,1st}/\Gamma_{\rm bulk,r} \sim$ 700 MeV, which is larger than the electron rest mass energy $m_{\rm e} c^2$. At these energies, the scattering cross-section is highly suppressed, and therefore we ignore the second-order IC emission for this GRB.
To bring the maximum SSC photon or spectral cutoff at GeV energies, the maximum accelerated electron energy ($\gamma_{\rm max, r}$) of the reverse-shock synchrotron component should be lower than $\gamma_{\rm c,r}$ (e.g., $\gamma_{\rm max, r} \sim$ 5 $\times$ 10$^3$ $\sim \,\,3 m_{\rm p}/m_{\rm e}$).

Thus, for GRB~180720B, we see possible evidence for a spectral cutoff, but this cutoff depends sensitively on the reverse shock acceleration mechanism, which originates from GRB ejecta itself characterized by the nature of the GRB central engine, e.g., the initial strong magnetic field around the black hole\cite{1997ApJ...482L..29M, 2009MNRAS.397.1153K}.

Previous works investigated the low synchrotron optical flux from the reverse shock in GRBs while assuming a high magnetization in the GRB ejecta\cite{2007MNRAS.378.1043J}.
As seen in this GRB, a high $Y$ value may suppress the synchrotron optical flux from the reverse shock without assuming a high magnetization. Thus, few previous works constrain the $Y$ value directly due to a lack of the simultaneous observation of the synchrotron and SSC components from the reverse shock and our finding gives important implications for understanding the reverse-shock physical mechanism.

\subsection*{Theoretical modeling of the afterglow with the time-independent parameters (``EATS model'')}
We also tested the theoretical modeling with the time-independent shock microphysical parameters to reproduce the afterglow emission 
in the early and late phases (Intervals II and III). This model\cite{Gill-Granot-18} calculates the observed flux at any given observer's 
time $T_{\rm obs}$ by integrating over the equal arrival time surface (EATS) of the GRB jet, which includes contributions from emission 
arising from shocked gas within the beaming cone (of angular size $1/\Gamma_{\rm bulk}$; see panel $a$ of Fig.\,\ref{fig:pol_model}) centered at the 
observer's line-of-sight (LOS). An EATS integration then 
properly accounts for the simultaneous arrival time of photons that were emitted at an earlier central-engine-frame time by material at small angular 
distance away from the LOS and those emitted by the shocked gas along the LOS but at a later central-engine-frame time. 
Thus, this model describes the realistic flux and spectral evolution of the afterglow emission.

By adopting the parameters as shown in Extended Data Table \ref{table:parameters_rs_fs} (see ``EATS''), the model can 
reproduce the observed multi-wavelength data at time intervals II and III with the time-independent parameters 
(Extended Data Fig. \ref{fig:EATS-model}). Note that $\gamma_{\rm m}$ and $\gamma_{\rm c}$ 
are not shown in Extended Data Table \ref{table:parameters_rs_fs} because those values are continuous for the EATS model 
and a single value cannot be easily defined.

 By reproducing the LAT lightcurve, this model clearly demonstrates that the origin of the LAT GeV flux is SSC emission from the reverse shock.
A jet-break is not added to the EATS model in order to reproduce the X-ray lightcurve in the late phase. This model deviates from the radio data and a similar difficulty to reproduce the radio observations was reported in the multi-wavelength study from the radio to TeV band\cite{2019Natur.575..459M}.
One of the important features of the EATS model is that the synchrotron emission from the reverse shock extends well into the X-ray band, which is different from the analytical model that only includes photons emitted along the line-of-sight (LOS). At the time of interval II, the X-ray band is above the cutoff frequency (above which the shocked plasma cannot radiate after reverse-shock crossing) for radiation emitted along the LOS. Therefore, flux contribution to the X-ray band only comes from radiation emitted from small angles away from the LOS and at earlier lab-frame times when the cutoff frequency was still above the X-ray band. 
For this reason, the detailed spectral modeling with EATS integration is important, and earlier analytical works may not have considered this effect.

This model gives a very low value of $\epsilon_{B,{\rm f}}$ ($\sim$10$^{-4}$) in the early phase, which induces the 
strong SSC component in the TeV band and may cause secondary cascade emission and contribute to the GeV band. To check 
if the TeV emission induces the secondary cascade emission, we calculate the opacity of the TeV emission. First, the 
TeV photons with $E_{\rm TeV} \sim  10^{12}$ eV mainly interact with target photons with 
$E_{\rm t} = (m_e c^2)^2 \Gamma_{\rm bulk}^2 E_{\rm TeV}^{-1}(1+z)^{-2} \sim$ 10 keV $(\Gamma_{\rm bulk}/300)^2 (E_{\rm TeV}/1 {\rm TeV})^{-1}$.
The target photon flux at $E_{\rm t} \sim$ 10  keV is roughly $F_{\rm t} \sim$ $ 5 \times $10$^{-9}$ erg/cm$^2$/s, 
i.e., the target photon luminosity $L_{\rm t}$ = 4 $\pi d_{\rm L}^2 F_{\rm t} \sim$  $10^{49}$ erg/s, where $d_{\rm L}$ 
is the luminosity distance of 4.0 Gpc. Thus, the target photon density $n_{\rm t}$ can be obtained as 
$L_{\rm t}/4\pi R^{2}\Gamma_{\rm bulk}cE_{\rm t}$, where $R$ is the radius of the GRB ejecta and 
$R \sim 4 c  \Gamma_{\rm bulk}^2 t_{\rm obs}(1+z)^{-1} \sim $10$^{18}$ cm, where $c$ is the speed of light.
Then, we calculate the opacity of the TeV emission in the GRB emission site to $\tau = n_{\rm t} \sigma_{\gamma\gamma} R/\Gamma_{\rm bulk}$ $\sim$ 10$^{-3}$, where  $\sigma_{\gamma\gamma}$ is the cross section of the $\gamma-\gamma$ annihilation being roughly  $\sim 0.1 \sigma_{\rm T}$ ($\sigma_{\rm T}$ is the Thomson cross section). This result indicates that the TeV emission is optically thin and the secondary cascade emission does not occur from the strong SSC photons.
 For the calculation of the model spectrum, only synchrotron photons from the forward shock region have been considered as target photons for IC scattering.
However, photons below 100 eV originated from the reverse shock region also contribute 
as IC seed photons, then the IC component should have a low-energy tail,
whose flux can potentially (requiring further more detailed investigation that is 
outside the scope of this work) exceed
the observed GeV flux in interval II. 
We note this weakness in such a high Compton $Y$ model in the early phase.

\subsection*{Do the early optical and GeV emissions arise from the tail of the prompt emission ?}
Fig. \ref{fig:LC} shows that the lightcurves at all frequencies appear to be decaying at a consistent rate for the first few hundred seconds and they may have the same origin, such as the tail of the prompt emission\cite{2005Natur.436..985T}.
One of the strong pieces of evidence for the tail of the prompt emission is the spectral softening with time, which was observed in the \Swift GRBs\cite{2007ApJ...666.1002Z} and can be interpreted as the curvature effect of a spherical, relativistic jet\cite{2000ApJ...541L..51K,2007ApJ...659.1420T}.
For the XRT and BAT data of this GRB, the significant spectral softening was observed as seen in the photon index evolution of Fig. \ref{fig:LC}, which infers that the early emission may arise from the tail of the prompt emission.
For some previous LAT GRBs, the spectral softening was  observed\cite{2019ApJ...886L..33A}, which can be also interpreted as the tail of the prompt emission.
 However, in the LAT lightcurve of this GRB, no such spectral softening was seen in the first few hundred seconds. 
 Instead, the photon index remains almost flat. 
 
For the optical data, the spectral evolution cannot be observed due to photometry in a single band. 
Here, if the early optical and GeV emissions  originate from the tail of the prompt emission, it should dominate the other component, namely, the observed SED can be represented by a single component such as a synchrotron model\cite{2019A&A...628A..59O}. However, the observed SED indicates that there are multiple components in Interval II. 
Note that although  the prompt optical emission for several GRBs has a different spectral component from  the X-ray and gamma-ray emission\cite{2008Natur.455..183R,2017ApJ...849...71L}, the observed temporal indices of the prompt optical emission (e.g., $\sim$-7 and $\sim$-14 for GRBs 080319B and 160625B, respectively) are much steeper than that of this GRB. 
 Thus, the observed spectral and temporal properties in the optical and GeV bands disfavor the scenario of the tail of the prompt emission for this GRB.

\section*{Data availability}
The Fermi-LAT data are publicly available at the Fermi Science Support Center website: {\small\url{https://fermi.gsfc.nasa.gov/ssc/}}. 
Swift XRT and BAT products are available from the online GRB repository {\small\url{https://www.swift.ac.uk/xrt_products}}.
All the raw data of HOWPol and HONIR can be downloaded at SMOKA data archiving site in NAOJ website: {\small\url{https://smoka.nao.ac.jp/index.jsp}}. The processed data are available from the corresponding author upon request.
%Filtered data supporting the findings of this study are available at this url: {\small\url{https://www-glast.stanford.edu/pub_data/1801/}}.

\section*{Code availability}
%In this study we use the Fermi-LAT Science Tools, publicly available at the Fermi Science Support Center website: \url{https://fermi.gsfc.nasa.gov/ssc/}. 
The details of the code are fully described in Methods.
The code to reproduce each figure of the paper is available from the corresponding author upon request.

%%%%%%%%%%%%%%%%%%%%%%%%%%%%%%%%%%%%%%%%%%%%%%%%%%%%%%%%%%%%%%%%%%%%%%%%%%%%%%%
%%%%%%% to count the word, uncomment the next line \end{document} and use the Overlef Menu-. Word Count
%\end{document}
%%%%%%%%%%%%%%%%%%%%%%%%%%%%%%%%%%%%%%%%%%%%%%%%%%%%%%%%%%%%%%%%%%%%%%%%%%%%%%%
\section*{Acknowledgements}
The \fermi-LAT Collaboration acknowledges support for LAT development, operation and data analysis from NASA and DOE (United States), CEA/Irfu and IN2P3/CNRS (France), ASI and INFN (Italy), MEXT, KEK, and JAXA (Japan), and the K.A.~Wallenberg Foundation, the Swedish Research Council and the National Space Board (Sweden). Science analysis support in the operations phase from INAF (Italy) and CNES (France) is also gratefully acknowledged. 
This work performed in part under DOE Contract DE-AC02-76SF00515 and support by MEXT/JSPS KAKENHI Grant Numbers JP17H06362 and JP23H04898, the JSPS Leading Initiative for Excellent Young Researchers program and the CHOZEN Project of Kanazawa University (M.A.). R.G. acknowledges financial support from the UNAM-DGAPA-PAPIIT IA105823 grant, Mexico. 
J. G. acknowledges financial support by the ISF-NSFC joint research program (grant no. 3296/19).

\section*{Author Contributions Statement}
The contact author for this paper is M. Arimoto who contributed to the analysis of X-ray and GeV data, the interpretation and the writing of the manuscript.
K. Asano., K. Toma, R. Gill, and J. Granot provided the interpretation and contributed to the writing of the paper. S. Razzaque contributed to the interpretation of the GRB model. K. Kawabata.,  K. Nakamura, T. Nakaoka, K. Takagi, M. Kawabata, M. Yamanaka and M. Sasada contributed to the optical Kanata observations and the optical data analysis. M. Ohno, S. Takahashi, N. Ogino, H. Goto analyzed X-ray and GeV data.
All authors reviewed the manuscript. 

%{\small{\url{arimoto@se.kanazawa-u.ac.jp} or \url{makoto.arimoto@gmail.com}}}.
%The authors declare no competing financial interests. All authors reviewed the manuscript. 
%Must include all authors, identified by initials, for example: A.A. conceived the experiment(s),  A.A. and B.A. conducted the experiment(s), C.A. and D.A. analysed the results.  All authors reviewed the manuscript. 

\vspace{1cm}
%\textbf{The \Fermi LAT and Kanata Collaboration}
%M.~Ajello$^{1}$, 
%W.~B.~Atwood$^{2}$, 
%M.~Axelsson$^{3,4}$, 

\section*{Competing Interests Statement}
The authors declare no competing interests.

\section*{Correspondence}
Correspondence and requests for materials should be addressed to M.A. ({\small{\url{arimoto@se.kanazawa-u.ac.jp}}}).

\medskip
%\begin{enumerate}
%\item[1.] Department of Physics and Astronomy, Clemson University, Kinard Lab of Physics, Clemson, SC 29634-0978, USA
%\item[2.] Santa Cruz Institute for Particle Physics, Department of Physics and Department of Astronomy and Astrophysics, University of California at Santa Cruz, Santa Cruz, CA 95064, USA
%\item[3.] Department of Physics, Stockholm University, AlbaNova, SE-106 91 Stockholm, Sweden
%\item[4.] Department of Physics, KTH Royal Institute of Technology, AlbaNova, SE-106 91 Stockholm, Sweden
%\item[*] Contact Author
%\end{enumerate}

\setcounter{figure}{0}

\captionsetup[figure]{format=plain,  name=Extended Data Fig.}

\captionsetup[table]{format=plain,  name=Extended Data Table}

\newpage

\begin{figure*}[ht!]   
\begin{center}
\includegraphics[width=1.1\textwidth, viewport=0 0 610 642]{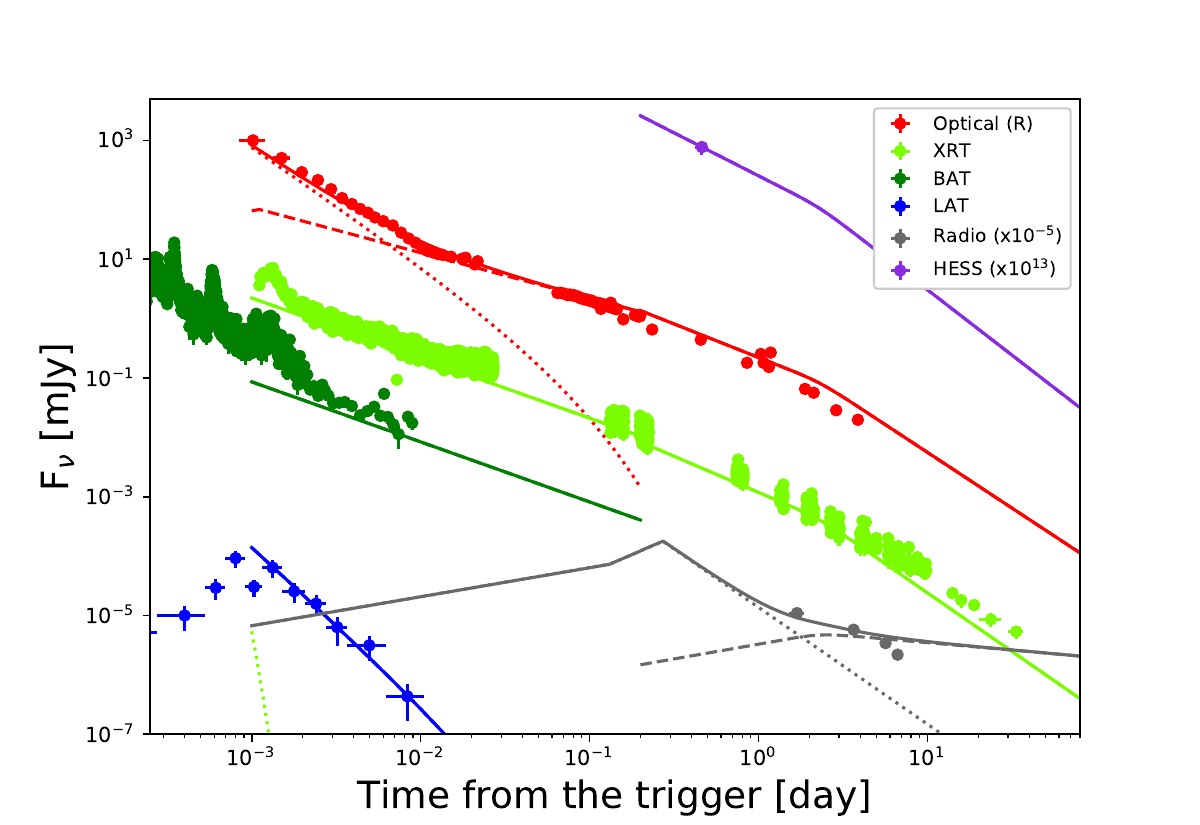}
\caption{
\textbf{Lightcurves of the afterglow with the analytical model.}
The observed flux density lightcurves at different frequencies (\Fermi-LAT at 300 MeV, HESS at 300 GeV, optical at 4.6 $\times$ 10$^{14}$ Hz, \Swift-XRT at 2 keV, \Swift-BAT at 30 keV, and radio at 15.5 GHz) are shown along with the theoretical reverse-shock (dotted), forward-shock (dashed) and combined reverse-shock plus forward-shock (solid) components. Note that the reverse-shock emission in the XRT band is suppressed because the maximum synchrotron frequency is much lower than the X-ray band. Errors correspond to the 1-$\sigma$ confidence region.
}
\label{fig:LC-model} 
\end{center}   
\end{figure*}

\clearpage

\begin{figure*}[ht!]   
\begin{center}
\hspace*{-17em} % move the table to left
\includegraphics[width=0.6\textwidth, viewport=0 0 710 1142]{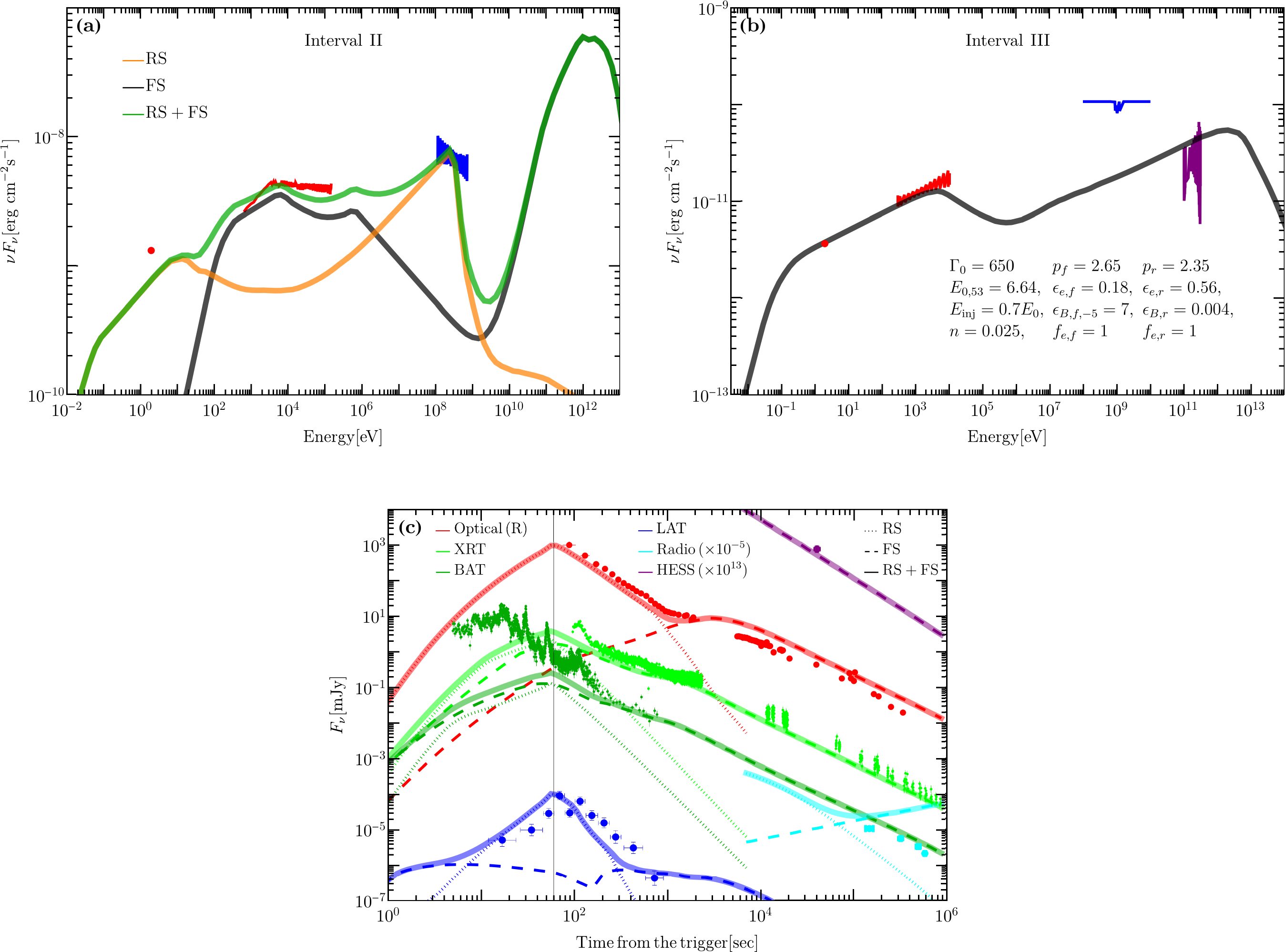}
\caption{\textbf{Theoretical model with time-independent parameters at time intervals II and III.}
({\bf a}) Spectral energy distribution at time interval II with the EATS model. The reverse- (RS) and forward-shock (FS) 
components are shown with the synchrotron and SSC emission. 
({\bf b}) Spectral energy distribution at time interval 
III with the EATS model. The legend shows the adopted model parameters. Here $\Gamma_0$ is the bulk Lorentz factor 
of the coasting flow before it is decelerated by the ISM, $E_{\rm inj}$ is the amount of energy injected during 
the shallow plateau phase, and subscripts `$f$' and `$r$' refer to FS and RS parameters, respectively. 
({\bf c}) Multi-waveband lightcurve and model comparison. The vertical line shows the duration of the prompt GRB. 
See the caption of Extended Data Fig.\,\ref{fig:LC-model} for details. Errors correspond to the 1-$\sigma$ confidence region.
}
\label{fig:EATS-model} 
\end{center}   
\end{figure*}

\begin{figure*}[ht!]   
\begin{center}
\includegraphics[width=0.8\textwidth, viewport=0 0 610 642]{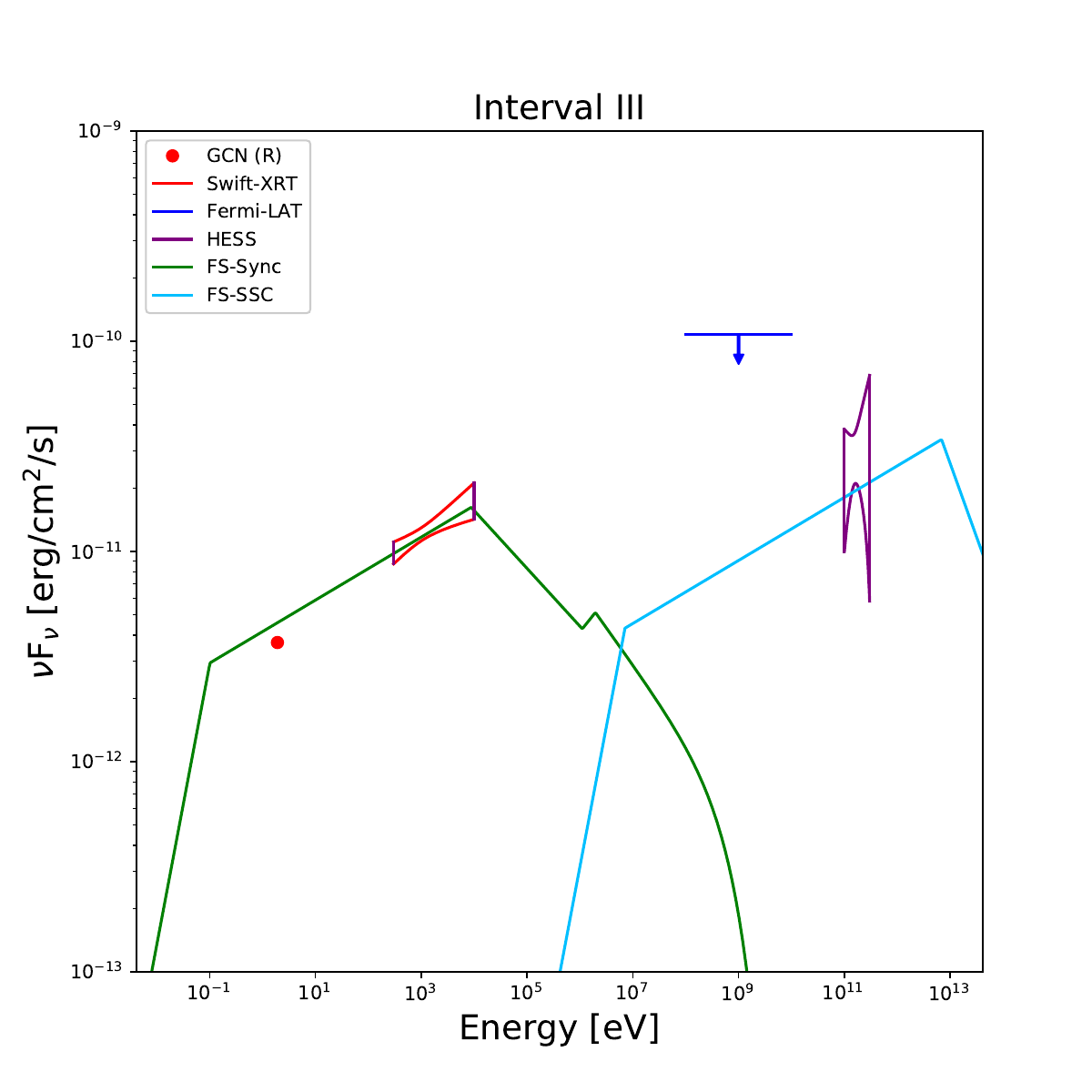}
\caption{\textbf{Spectral energy distribution at time interval III.}
The solid lines in the low-energy and high-energy bands represent  the synchrotron and SSC components from the forward shock with the ``analytical'' model, respectively.
The red area corresponds to the 1-$\sigma$ confidence region from the best-fit power-law function   for the \Swift-XRT. Note that the XRT observation was not actually performed in the time interval and we used the interpolated flux before and after the interval (this interpolation is reasonable because the photon index is almost constant from $T_0$ + 10$^4$ s to 10$^5$ s, as shown in the bottom panel of Fig. \ref{fig:LC}). The blue arrow represents the 90\% upper limit in the \Fermi-LAT range.  The red  point represents the optical flux observed by the optical telescope.
The purple area represents the 1-$\sigma$ confidence region from the best-fit power-law function for the HESS. 
}
\label{fig:SED_int3} 
\end{center}   
\end{figure*}

\clearpage

\begin{table*}[hb]
\begin{center}
\caption{Model parameters used for reverse and forward shocks afterglow modeling and output parameters. }\label{table:parameters_rs_fs}
\hspace*{-3em} % move the table to left
\includegraphics[width=1.2\textwidth, viewport=0 0 610 542]{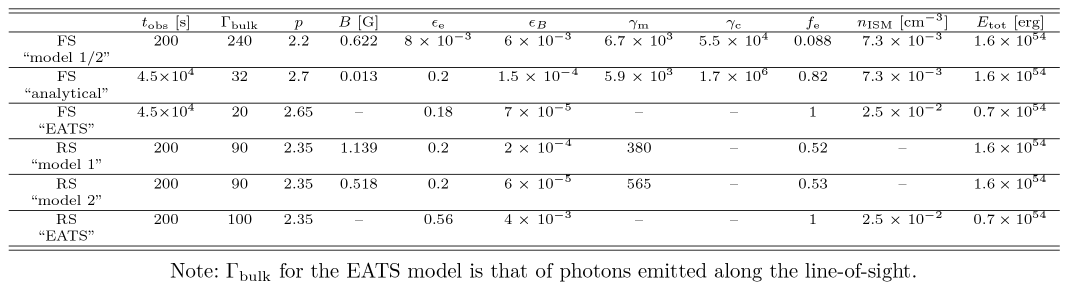}
\end{center}
%% Any table notes must follow the \end{tabular} command.
%\tablenotetext{a}{Sample footnote for table~\ref{tbl-2} that wasgenerated with the \LaTeX\ table environment}
%\tablenotetext{b}{Yet another sample footnote for table~\ref{tbl-2}}
%\tablenotetext{c}{Another sample footnote for table~\ref{tbl-2}}
%\leavevmode \\{
%Note: $\Gamma_{\rm bulk}$ for the 
%EATS model is that of photons emitted along the line-of-sight.
%}
\end{table*}

\clearpage

\section*{Supplementary Methods}

\subsection*{Statistical significance of the GeV excess}
To estimate statistical significance of  the GeV excess over the X-ray extrapolation in Interval II, we used the following models: a broken power-law function, a broken power-law function + a single power-law function. Here, we assume that a single power-law  function corresponds to the GeV component. We fitted the X-ray and GeV spectra with the two models and the obtained best-fit parameters are shown in Supplementary Table \ref{table:spectrum_XRT_BAT}. Note that the photon index of the single power-law function is fixed because the photon index is not well determined.  By incorporating the single power-law function,  the statistic value $S$ is improved by $\Delta S$ = 14 for a decrease of the number of degrees of freedom by one, where $\Delta S$ is defined as the difference between statistic values of the two models. If $S$ follows the $\chi^2$ distribution for Wilks theorem, we may analytically derive the null hypothesis probability. However, it is not obvious that $S$ follows the $\chi^2$ distribution and then we perform the Monte Carlo simulation. 
First, we generated 2 $\times$ 10$^5$ fake spectra randomly assuming the broken power-law function as the null hypothesis model. Then, we fitted each of the generated spectrum with the null hypothesis model and  the alternative model (i.e., a broken power-law function + a single power-law function). The difference $\Delta S$ between the null hypothesis and alternative models can be obtained for each realized spectrum. Finally, the complementary cumulative distribution function as a function of $\Delta S$ is obtained as shown in Supplementary Fig. \ref{fig:Prob_GeVexcess}.
The result shows that $\Delta S = 14$ corresponds to the null hypothesis probability being $\sim$6 $\times$ 10$^{-5}$, which indicates that the GeV excess exists with 4.1 $\sigma$ confidence level.

\subsection*{Optical polarimetric observations and results}
We performed optical (R$_c$-band) polarimetric observations of GRB~180720B with HOWPol\cite{2008SPIE.7014E..4LK} and HONIR\cite{2014SPIE.9147E..4OA} for $T_0$ + 73~s to 1880~s and $T_0$ + 4620~s to 16720~s, respectively. HOWPol and HONIR were attached to the Nasmyth and the Cassegrain foci, respectively, of the 1.5-m Kanata telescope at Higashi-Hiroshima Observatory.
Our observation with HOWPol was automatically processed after receiving the {\it Swift}/BAT Notice via GCN. 
With HOWPol, we took ten 30-s exposures and then twenty 60-s exposures in a wide-field (without focal mask), one-shot polarimetry mode through a wedged double Wollaston prism, providing four linearly polarized images at instrumental position angles (PAs) of 0$^\circ$, 90$^\circ$, 45$^\circ$ and 135$^\circ$. 
This enables us to obtain all three Stokes parameters for linear polarization, i.e., $I, Q, U,$ from a single exposure.
Examples of the raw images acquired by HOWPol and HONIR are shown in Supplementary Fig. \ref{fig:GRB_opt_image}.
We calculated the Stokes parameters in the same way as a traditional polarimetry through a polarizing filter rotated by 45$^\circ$ step.
With HONIR, we took both Rc-band image (155 s exposure) and H-band image
(140 s exposure) simultaneously in a dual-beam (ordinary and extraordinary rays) polarimetry mode, through a rotating superachromatic half-wave plate and a fixed Wollaston prism with a focal mask avoiding a superimposition of ordinary and extraordinary rays. Each observation unit consisted of a sequence of exposures at four PAs of the half-wave plate, 0$^\circ$, 45$^\circ$, 22.5$^\circ$ and 67.5$^\circ$. We calculated the Stokes parameters as described in the HONIR reference paper\cite{2014SPIE.9147E..4OA}.

For subtraction of the instrumental polarization, correction of the instrumental depolarization and the conversion of the coordinate (instrumental to celestial), we used observations of unpolarized and polarized standard stars\cite{serkowski1975wavelength,schmidt1992hubble}, including measurements through a fully polarizing wire grid, taken in the same observation period. 
The polarimetry with HOWPol suffers from large instrumental polarization produced by the reflection on the tertiary mirror of the telescope. The instrumental polarization was modeled as a function of the declination of the object and the hour angle at the observation.
The instrumental polarization (PD$_{\rm Rc,HOWPol}\simeq 3.7$\%\ for HOWPol and PD$_{\rm Rc,HONIR}\leq 0.1$\% for HONIR) was vectorially removed in each data point. The uncertainty of the modeled instrumental polarization gives additional systematic error of $\Delta PD\simeq 0.5$\%\ in HOWPol data.
The depolarization factor is negligibly small ($\leq 0.01$) for both HOWPol and HONIR polarimetry, and we did no correction for it.
Using the polarimetric results of polarized standard stars, we converted the PA of the polarization into the celestial equatorial coordinate. Besides, we confirmed the fully calibrated polarization data are consistent with those in the literature.

The observed polarization of GRB~180720B and  nearby comparison stars (C1 and C2) are shown in 
Fig. \ref{fig:QU}.
C1 and C2 are stars with the Rc-band magnitude of $R_{c, C1}=12.24\pm 0.03$ and $R_{c, C2}=11.99\pm 0.03$ located 0.8-arcmin south and 2.4-arcmin northeast from GRB~180720B, respectively.  It is noted that C1 and C2 were chosen just as the polarimetric comparison stars of which the brightness is comparable to the GRB afterglow in the early phase and that neither C1 or C2 was used for the polarimetric calibration described above. 
Before $T_0$ + 450 s the afterglow of the GRB is brighter than C1 or C2. 
The very bright optical afterglow enabled us to measure its polarization very precisely in the early phase of the GRB emission. 
The main error budget of each data point is the combined photometric errors (e.g., photon statistical uncertainties of the object and background sky) of the four or eight images for a single object. 
In the HOWPol observation, the throughput of the images polarized at instrumental PAs of $45^\circ$ and 135$^\circ$ was worse than that at $0^\circ$ and 90$^\circ$. 
This gives asymmetric uncertainties among $Q/I$ and $U/I$, which gives  larger uncertainties of the $Q/I$ ($\sim$2\%) than those of the $U/I$ ($\sim$0.5\%) for the C1 and C2 stars in the first 10 images taken with a short exposure (30 s). 

We find that there is no correlation between the observed polarization of GRB~180720B and C1/C2 as shown in the $Q/I-U/I$ diagram (Supplementary Fig. \ref{fig:QU_time_series}).
Thus, for the time interval of $T_0$ + 70 -- 300 s (the left part of Fig. \ref{fig:QU}), we see a temporal evolution of the GRB PD intrinsically from 5\% to  $\lesssim$1\% with the almost stable PA between $\sim$50$^\circ$ and $\sim$80$^\circ$.
For the time interval of $T_0$ + 300 -- 1000 s (the middle part of Fig. \ref{fig:QU}), the optical GRB flux is comparable with or lower than C1 or C2. Thus, the uncertainties of $Q/I$ and $U/I$ of GRB~180720B are of the same order of magnitude as C1 ($\sim$1--2\%). As shown in the $Q-U$ diagram of the average of C1 and C2 (Supplementary Fig. \ref{fig:QU_time_series}), the polarimetric data points are concentrated around $U/I \sim Q/I \sim$ 0\% . 
In contrast, the data points of GRB~180720B have significantly a different distribution from that of the average of C1 and C2, which indicates that intrinsic polarization of the GRB emission was detected.

For the time interval of $T_0$ + 1000 -- 2000 s, the GRB emission becomes much fainter than C1 or C2 and the signal-to-noise ratio becomes significantly worse for the HOWPol observation  (Supplementary Fig. \ref{fig:QU_time_series}). Thus, the uncertainties of $Q/I$ and $U/I$ of GRB~180720B are large and the polarization vectors of GRB~180720B are  not well determined. 
Note that as shown in the $Q/I-U/I$ diagram (Supplementary Fig. \ref{fig:QU_time_series}) the polarization vectors of GRB~180720B at $T_0$ + 1500 -- 2000 s are likely distributed in the opposite directions of those at  $T_0$ + 300 -- 1000 s, which may indicate that the PA changes from $\sim$100$^\circ$ at $T_0$ + $\sim$1000 s to PA $\sim$ 0$^\circ$ (= 180$^\circ$) at $T_0$ + $\sim1800$ s. 
The time at $T_0$ + $\sim$1000 s correspond to a transition phase from reverse shock to forward shock. The polarization vectors at $T_0$ + 1500 -- 2000 s are roughly consistent with those at $T_0$ + 5000 s or later.
 
At $t_{\rm obs} \gtrsim T_0$ + 5 $\times$ 10$^3$$\;$s, we performed optical polarimetry in a more accurate way as described above. As shown in the right part of 
Fig. \ref{fig:QU}, the uncertainty of PD of the average of C1 and C2 in each data point is small to be $<$0.3\% (the error bar is less than the marker size) and the polarization was precisely measured. 
The average polarization of C1 and C2 is obtained to be 0.25\%$\pm$0.03\% with PA = 120.5$^\circ$ $\pm$ 5.6$^\circ$ (this measured value was confirmed a few months after the burst occurrence). 
Then, we regard the measured polarization as the interstellar polarization (ISP). Note that the measured polarization vectors of GRB~180720B are corrected by subtracting the ISP contribution for both the HOWPol and HONIR data points.
Due to the superior polarimetric observation with HONIR, the PD and PA of the GRB afterglow are well determined, albeit the low polarization degree of GRB~180720B (PD $\sim$ 1--2\%). The PA observed by HONIR ranges from $\sim$150$^\circ$ to $\sim$190$^\circ$ (= $\sim$10$^\circ$), i.e., $\Delta$PA $\sim$ 40$^\circ$. The variation of the PA seems less significant than that in the earlier phase of $T_0$ + 70 -- 1000 s ($\Delta$PA $\sim$ 90$^\circ$) where the reverse shock is dominant. These features would indicate that the polarization behavior of the forward shock is different from that of the reverse shock.

\subsection*{Prompt emission and inefficient particle acceleration}
Three instruments on board the two space observatories --  
\Fermi(GBM and LAT) and \Swift-BAT, covering the energy range from 8 keV to 100 GeV, detected intense $\gamma$-rays from the prompt emission phase of GRB~180720B. The duration of the prompt 
emission, as detected by \Fermi-GBM, is $\sim$60$\;$s, where the emission shows strong temporal variability that 
likely originated in internal shocks\cite{1999PhR...314..575P}.
This burst is very bright and its 10$\;$--$\,$10$^4\;$keV fluence is 
3$\,\times\,$10$^{-4}\;$erg/cm$^2$, which is in the top $<$1 percentile of GRBs in the GBM catalog\cite{2020ApJ...893...46V}. At a redshift of $z=0.654$\cite{2018GCN.22996....1V} the time-integrated isotropic equivalent energy 
of this GRB is $E_{\rm iso}=5\times10^{53}\;$erg (1$\,$--$\,$10$^4\;$keV). 
In addition, few $>$1$\;$GeV photons are detected during the prompt emission phase 
The highest energy photon observed by \Fermi-LAT is a 5$\;$GeV event at $T_0+142\;$s, which arrives  after the prompt emission ends.

The  composite lightcurves observed by \Fermi-GBM, \Fermi-LAT and \Swift-BAT in the prompt emission are shown in Supplementary Fig. \ref{fig:CompositeLC_prompt}. For most of the time intervals of the prompt emission phase, the observed spectra are mainly consistent with  the smoothly connected broken power-law functions with one or two breaks (i.e., the broken power-law function with one break corresponds to the Band function\cite{1993ApJ...413..281B}), which was also reported by previous work\cite{2020A&A...636A..55R}. In this paper, we focus on the brightest part of the prompt emission at  $T_0$ + 15.5  -- 17.5 s and the obtained spectrum and the corresponding best-fit parameters are shown in Supplementary Fig. \ref{fig:Spectrum_prompt} and Supplementary Table \ref{table:spectrum_prompt}, respectively.
The derived spectrum is represented by a smoothly broken power-law function with two spectral breaks and an exponential cutoff. The best-fit two photon indices indicate that the observed spectrum is almost consistent with either  the fast or slow cooling scenario. Due to the statistical uncertainty, we do not constrain which cooling scenario is favored.

The spectral cutoff energy $E_{\rm cut}$ is $\sim$26 MeV in the observer frame. If this spectral cutoff is caused by $\gamma$-$\gamma$ opacity, the cutoff energy should be $m_e c^2$ or more in the jet-comoving frame, namely, $E_{\rm cut}$ $>$ 0.511 MeV (1+$z$)$^{-1}  \Gamma_{\rm bulk}$, which leads to $\Gamma_{\rm bulk} < $ 84. Here,  the observed afterglow onset of $\sim$100 s corresponds to the deceleration time $t_{\rm dec}$.
By using Equation \ref{eq:dec_time}, 
we obtain $\Gamma_{\rm bulk}$ = 310. Thus, we find that the spectral cutoff energy in the jet comoving frame, $E^\prime_{\rm cut} \sim 140 \, {\rm keV} (\Gamma_{\rm bulk}/310)^{-1}$ is lower than  $m_e c^2$, which may indicate that the spectral cutoff is not caused by $\gamma$-$\gamma$ opacity and attributed to the intrinsic GRB synchrotron spectrum. 
If this is the case, the particle acceleration to the maximum energy in the prompt emission is inefficient. If  the acceleration timescale $t_{\rm acc}$ of the highest-energy electrons is determined by the Larmor radius 
$r_{\rm L}$ = $ \gamma m_e c^2/eB$  and the Bohm parameter $\xi$ (where $\xi \sim 1$  is the Bohm limit), $t_{\rm acc} = \xi 2 \pi \gamma m_e c / eB$. 
By equating $t_{\rm acc}$ with the synchrotron cooling timescale $t_{\rm sync} = 6\pi m_e c/\sigma_{\rm T} \gamma B^2$, the critical  synchrotron photon energy in the jet-comoving frame is $h\nu_{\rm max} = 9/8\alpha_f \xi$ = 80 $\xi^{-1}$ MeV\cite{2012MNRAS.427L..40K}, where $\alpha_f$ is the fine structure constant. Considering the estimated highest synchrotron energy of $140 \, {\rm keV} (\Gamma_{\rm bulk}/310)^{-1}$,  
the critical  synchrotron photon energy of this burst in the prompt emission is $h\nu_{\rm max} \sim$ 140 ($\xi$/570)$^{-1}$ keV, which suggests that the synchrotron emission from the internal shock does not reach the Bohm limit ($\xi \sim$ 1).

\clearpage

\captionsetup[figure]{format=plain,  name=Supplementary Fig}

\captionsetup[table]{format=plain,  name=Supplementary Table}

\setcounter{figure}{0}
\setcounter{table}{0}

\begin{figure*}[ht!]   
\begin{center}
\includegraphics[width=0.9\textwidth, viewport=0 0 610 642]{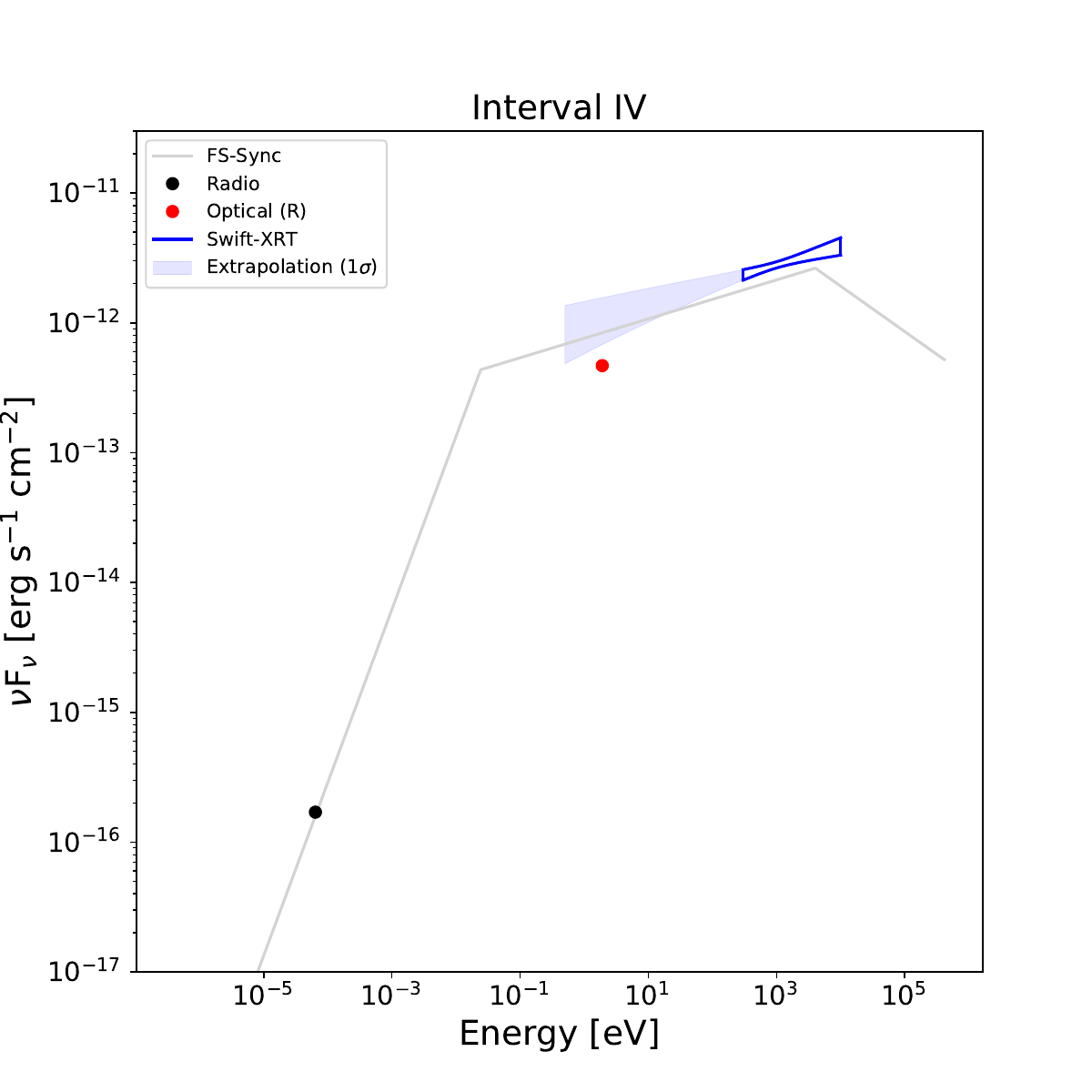}
\caption{\textbf{Spectral energy distribution at time interval IV. }
The blue area corresponds to the 1-$\sigma$ confidence regions from the best-fit power-law function   for the \Swift-XRT and \Fermi-LAT ranges, respectively.  The red and black  points represent the optical and radio fluxes observed by the Kanata telescope and AMI-LA.
The solid line represents the synchrotron spectrum of the forward shock afterglow emission.
}
\label{fig:SED_int4} 
\end{center}   
\end{figure*}

\begin{figure*}[ht!]   
\begin{center}
\hspace*{0em} % move the table to left
\includegraphics[width=1.1\textwidth, viewport=0 0 610 642]{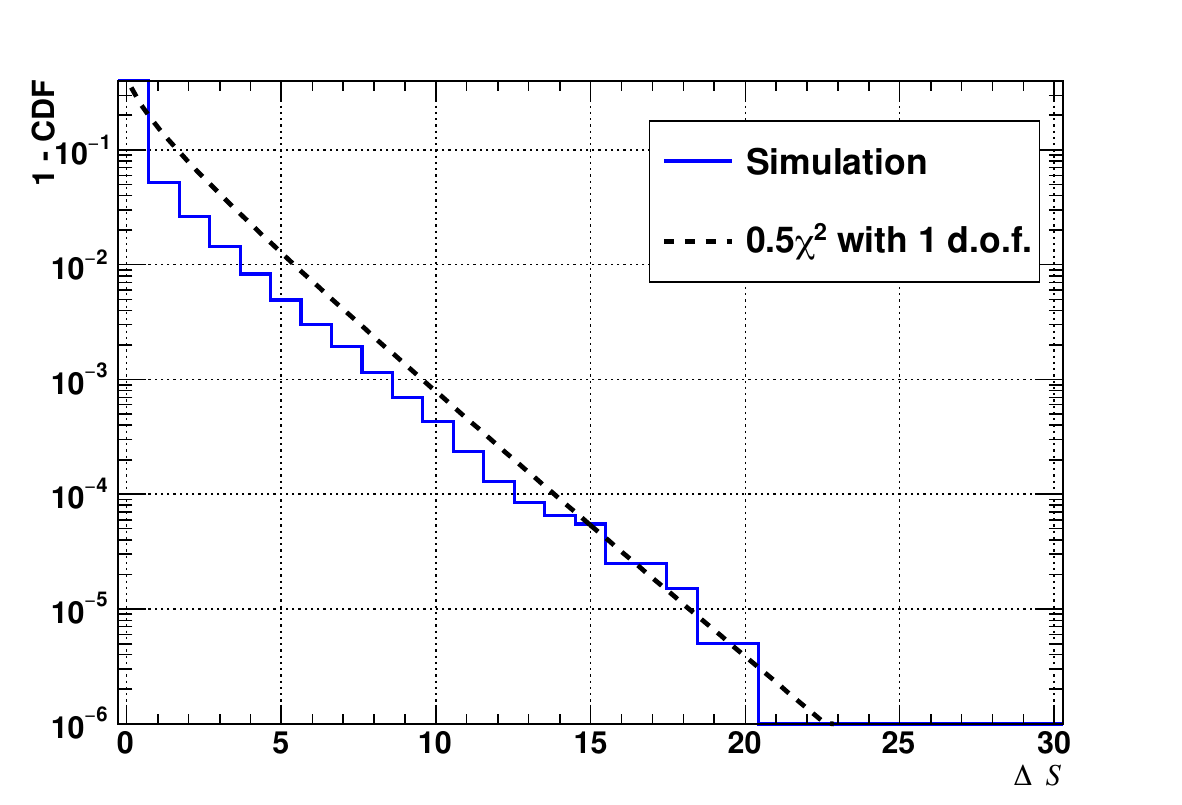}
\caption{\textbf{ Complementary cumulative distribution function (CDF) for $\Delta S$ between the null hypothesis and alternative models.}
The null hypothesis and alternative models correspond to a broken power-law function  and a broken power-law function plus a power-law function, respectively.
$\Delta S$ is difference between test statistics of the null hypothesis and alternative models.
The dashed line represents the 1 - CDF of $\chi^2$/2 with 1 d.o.f  as a reference probability distribution.
}
\label{fig:Prob_GeVexcess} 
\end{center}   
\end{figure*}

\clearpage

\begin{figure*}[ht!]   
\begin{center}
\hspace*{-6em} % move the table to left
\includegraphics[width=0.8\textwidth, viewport=0 0 610 642]{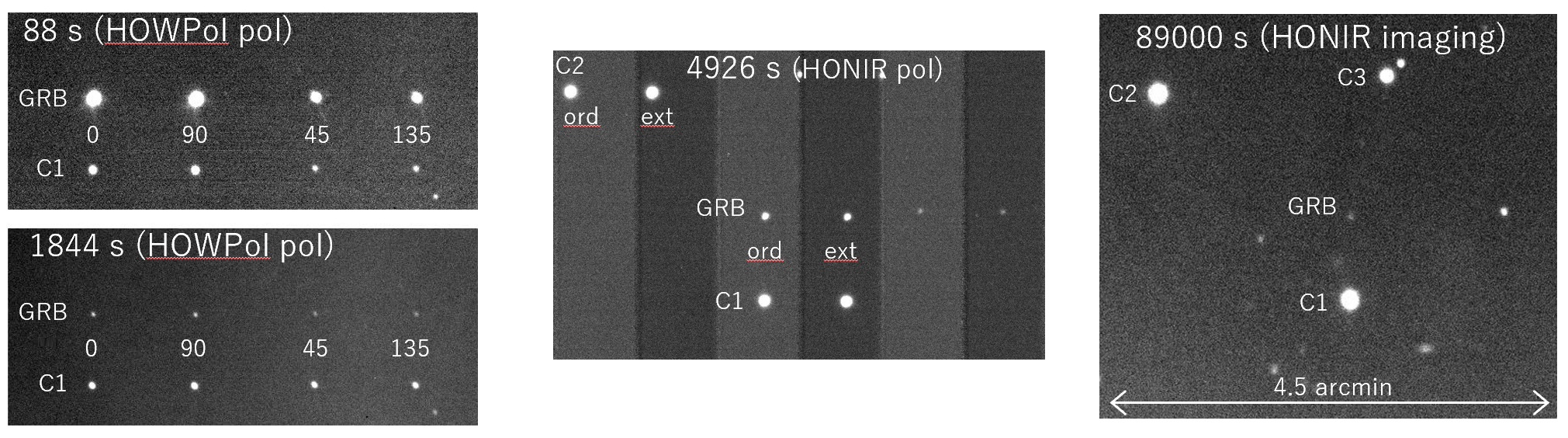}
\caption{\textbf{Optical raw images obtained by HOWPol and HONIR. }
The left panels show the obtained images with one-shot polarimetry mode of HOWPol. With a wedged double Wollaston prism, an image with different PAs can be obtained simultaneously (0$^\circ$, 90$^\circ$, 45$^\circ$ and 135$^\circ$). The middle panel shows an image for the polarimetric observation of HONIR with a dual-beam (ordinary and extraordinary rays) polarimetry mode. The ordinary and extraordinary rays can be obtained in a single image. The right panel shows an image with the imaging-mode observation.
}
\label{fig:GRB_opt_image} 
\end{center}   
\end{figure*}

\clearpage

\begin{figure*}[ht!]   
\begin{center}
\hspace*{-14em} % move the table to left
\includegraphics[width=0.8\textwidth, viewport=0 0 610 642]{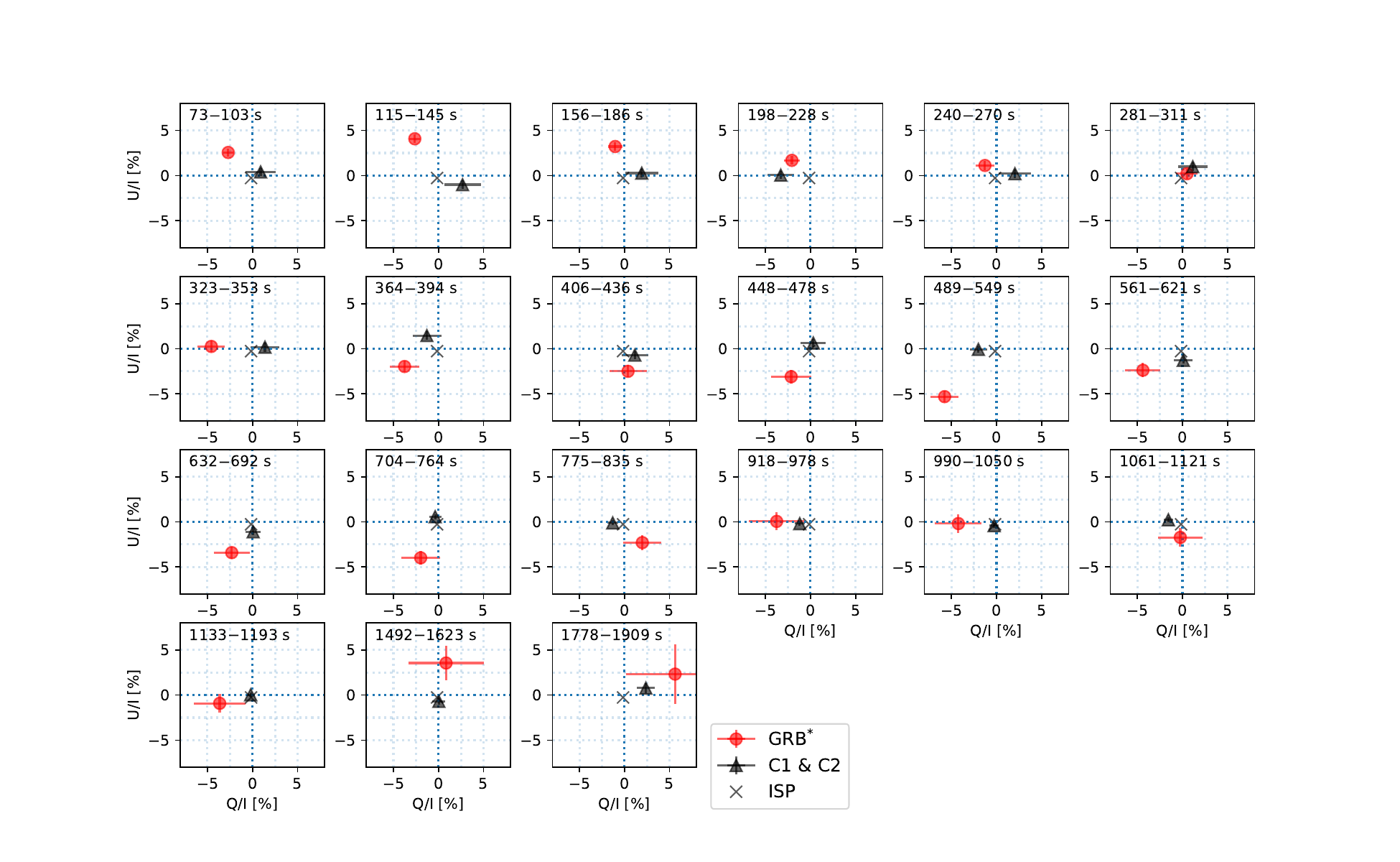}
\caption{\textbf{Time series plots of the Stokes parameters of GRB~180720B and the nearby stars. }
The time series plots of the $Q/I-U/I$ diagrams of the GRB and the average of the C1 and C2 stars  observed by HOWPol.
A ``$*$'' symbol indicates the the intrinsic GRB polarization after the subtraction of the interstellar polarization (ISP), 
where the ISP was precisely measured with the long-term measurement with HONIR. Note that the error bar of the ISP is much smaller than the marker size and all error bars correspond to the 1$\sigma$ confidence region. 
}
\label{fig:QU_time_series} 
\end{center}   
\end{figure*}

%\begin{figure*}[ht!]   
%\begin{center}
%\includegraphics[width=0.9\textwidth,trim=0 0 0 0,clip]{Figure/LC_radio_model.png}
%\caption{\textbf{Radio lightcurves with the theoretical forward shock synchrotron emission.}
%The red and blue points represent the observed radio data at 1.4 GHZ and 15.5 GHz, respectively, and the corresponding solid lines represent the theoretical curves from the adopted parameters of the forward-shock synchrotron model. 
%}
%\label{fig:LC_radio_model} 
%\end{center}   
%\end{figure*}

%\begin{figure*}[ht!]   
%\begin{center}
%\includegraphics[width=0.9\textwidth,trim=0 0 0 0,clip]{Figure/LC_GeV_model.png}
%\caption{\textbf{GeV lightcurve with the theoretical forward shock synchrotron emission.}
%The  blue points represents the observed GeV fluxes in the \Fermi-LAT band.  The corresponding solid line represents the theoretical curve from the adopted parameters of the forward-shock synchrotron model and the observed GeV flux cannot be fully explained by by roughly two orders of magnitude at $T_0$ + $\sim10^{-3}$ day ($\sim$100 s).
%}
%\label{fig:LC_GeV_model} 
%\end{center}   
%\end{figure*}

\clearpage

\begin{figure*}[ht!]   
\begin{center}
\hspace*{0em} % move the table to left
\includegraphics[width=1.0\textwidth, viewport=0 0 610 642,angle=0]{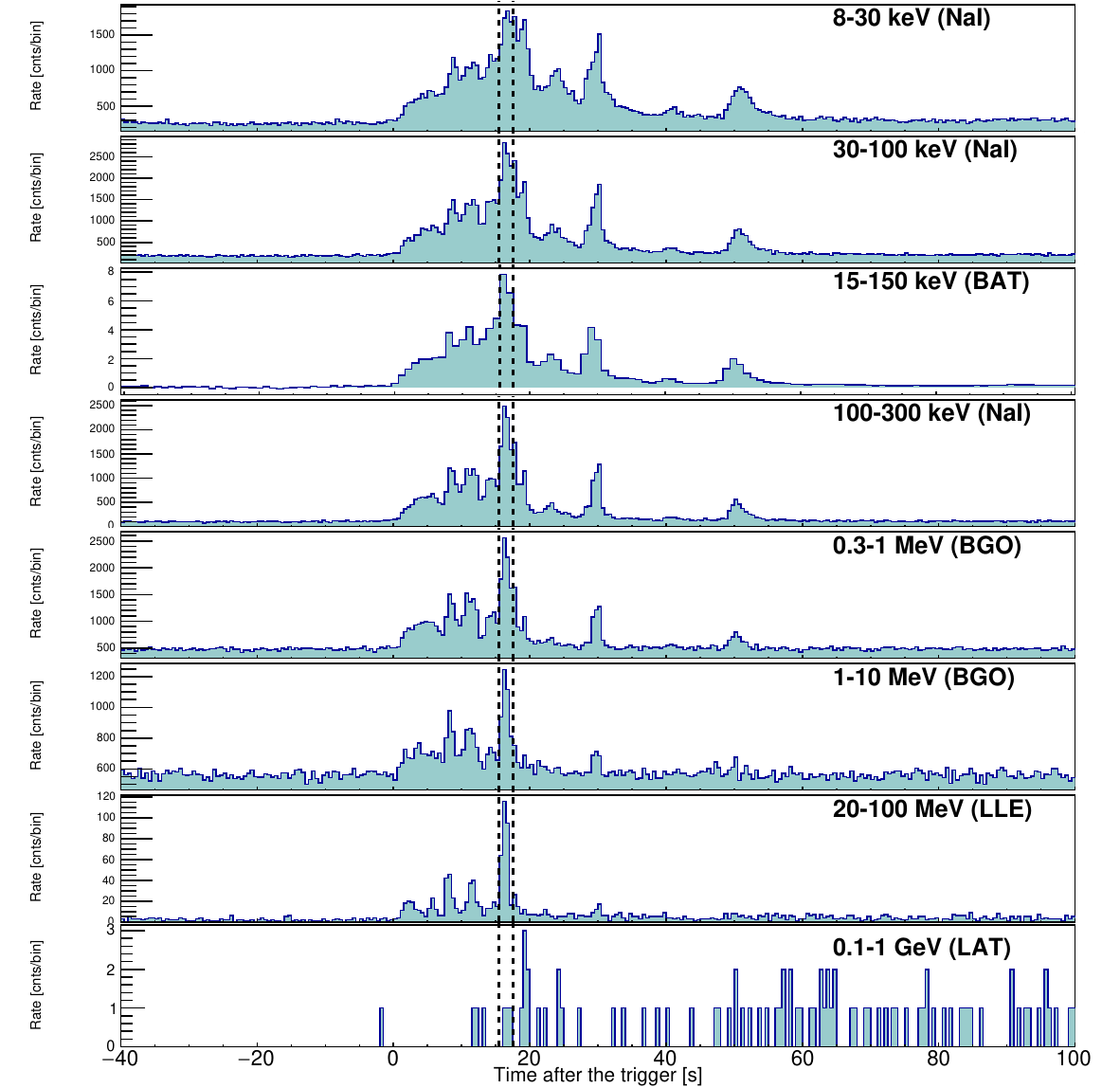}
\caption{\textbf{Composite lightcurve in the prompt emission phase.}
From the top, 8--30 keV (NaI), 30--100 keV (NaI), 15--150 keV (\Swift-BAT), 100--300 keV (NaI), 0.3--1 MeV (BGO), 1--10 MeV (BGO), 20--100 MeV (LLE) and 0.1--1 GeV (LAT),  where ``NaI'' and ``BGO'' are the detectors of \Fermi-GBM and ``LAT'' is the detector of \Fermi-LAT and ``LLE'' is the LAT low-energy data\cite{2010arXiv1002.2617P}. The time interval enclosed by the two dashed vertical lines ($T_0$ + 15.5  -- 17.5 s)  is the brightest emission part of this burst and the derived spectrum is shown in Supplementary Fig. \ref{fig:Spectrum_prompt}.
}
\label{fig:CompositeLC_prompt} 
\end{center}   
\end{figure*}

\clearpage

\begin{figure*}[ht!]   
\begin{center}
\hspace*{-5em} % move the table to left
\includegraphics[width=0.9\textwidth,viewport=0 0 610 642,angle=-90]{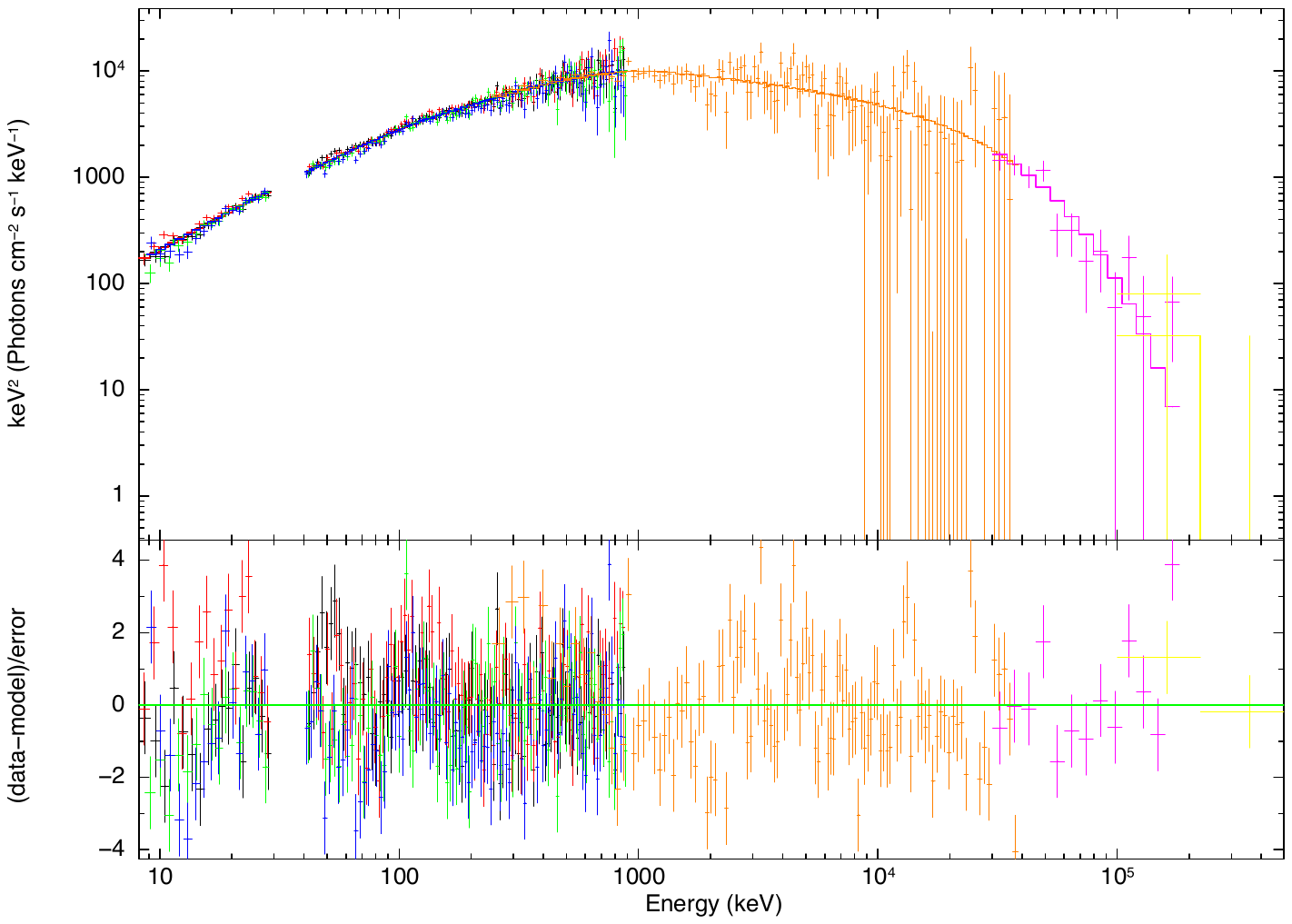}
\caption{\textbf{Prompt-emission spectrum at the brightest emission part.}
The observed spectrum is selected from $T_0$ + 15.5  -- 17.5 s.
It consists of 8--900 keV (NaI n6, n7, n8 and nb) and 230 keV--40 MeV (BGO b1), 30--200 MeV (LLE) and 0.1-1 GeV (LAT). Note that the energies of 30--40 keV was ignored to remove the non-astrophysical X-ray line of iodine from the NaI detector.
The best-fit function is represented by the smoothly connected broken power-law function with two breaks and the exponential cutoff. Errors correspond to the 1$\sigma$ confidence region.
}
\label{fig:Spectrum_prompt} 
\end{center}   
\end{figure*}

\clearpage

\begin{table*}[hb]
\begin{center}
\caption{Fitting result of the spectra in time intervals I, II and IV \label{table:spectrum_XRT_BAT}}
\hspace*{-6em} % move the table to left
%\small % added because Table spreads out of a page.
\footnotesize % added because Table spreads out of a page.
\begin{tabular}{cccccccc}
\hline \hline
%\begin{center}

 Interval & Model (instruments) & $n_H$  &$\Gamma_{\rm ph,1}$ & $E_{\rm break}$, &$\Gamma_{\rm ph,2}$  & $\Gamma_{\rm ph,ext}$ &  $S$ / d.o.f.  %\tablenotemark{a}  
\\
 &   &  [10$^{22}$ cm$^{-2}$] & &  $E_{\rm peak}$ [keV] &  &  &    %\tablenotemark{a}  
\\
%\end{center}
\hline
% total
I & pow (XRT+BAT) & 0.64$^{+0.09}_{-0.07}$ &-1.96$^{+0.02}_{-0.02}$ &  -- & --   & -- &   261/182   \\
80 -- 130 s & bknpow (XRT+BAT) &0.22$^{+0.11}_{-0.10}$ & -1.36$^{+0.17}_{-0.16}$ &  2.55$^{+0.57}_{-0.32}$ & -2.04$\pm$0.03& -- &  207/180   \\ 
& Band (XRT+BAT) &0.11$^{+0.09}_{-0.10}$ & -0.77$^{+0.36}_{-0.20}$ &  3.45$^{+0.76}_{-0.72}$ & -2.04$^{+0.03}_{-0.02}$& --  & 206/180   \\ \hline

II & pow (XRT+BAT)  & 0.32$^{+0.03}_{-0.03}$ &-1.91$^{+0.02}_{-0.02}$ &  -- & -- & --  & 392/353   \\
160 -- 300 s  & bknpow  (XRT+BAT)  &0.20$^{+0.04}_{-0.04}$ & -1.68$^{+0.07}_{-0.07}$ &  3.20$^{+0.50}_{-0.30}$ & -2.03$^{+0.05}_{-0.04}$ & -- & 330/351   \\ 
& Band  (XRT+BAT) &0.19$^{+0.04}_{-0.06}$ & -1.60$^{+0.23}_{-0.11}$ &  6.49$^{+4.28}_{-2.57}$ & -2.08$^{+0.08}_{-0.14}$ & -- & 331/351   \\ 
 & bknpow (XRT+BAT+LAT) & 0.20$^{+0.04}_{-0.04}$ & -1.69$^{+0.07}_{-0.06}$ &  2.87$^{+0.62}_{-0.52}$ & -1.97$^{+0.02}_{-0.02}$ & -- & 344/354   \\ 
  & bknpow + pow (XRT+BAT+LAT) & 0.20$^{+0.04}_{-0.04}$ & -1.68$^{+0.07}_{-0.07}$ &  3.23$^{+0.89}_{-0.48}$ & -2.07$^{+0.06}_{-0.05}$ & -1.7$^{\rm fixed}$ & 330/354   \\ 
\hline

%II & pow  & 0.33$^{+0.03}_{-0.03}$ &-1.89$^{+0.02}_{-0.02}$ &  -- & -- &   477/384   \\
%& bknpow  &0.22$^{+0.04}_{-0.04}$ & -1.68$^{+0.06}_{-0.07}$ &  3.02$^{+0.38}_{-0.52}$ & -2.06$^{+0.05}_{-0.05}$ &  416/382   \\ \hline

IV & pow (XRT) & 0.33$^{+0.09}_{-0.09}$ &-1.85$^{+0.10}_{-0.10}$ &  -- & -- & --  & 384/409   \\
(1.2 -- 1.8)$\times$10$^5$ s &  &  & &   &  &   &    \\

\hline
\end{tabular}
%% Any table notes must follow the \end{tabular} command.
%\tablenotetext{a}{Sample footnote for table~\ref{tbl-2} that wasgenerated with the \LaTeX\ table environment}
%\tablenotetext{b}{Yet another sample footnote for table~\ref{tbl-2}}
%\tablenotetext{c}{Another sample footnote for table~\ref{tbl-2}}
\end{center}
\leavevmode \\{``pow'', ``bknpow'', ``Band'' represent a power-law and  a broken power-law and  the Band functions, respectively. \\
The Galactic extinction is included as {\it tbabs} and the host extinction $n_H$ is calculated by using {\it ztbabs}, which models are implemented in the Xspec tool. \\
$E_{\rm break}$ and $E_{\rm peak}$ are the break energy of the broken power-law function and the peak energy  of the Band function in the $\nu F_\nu$ spectrum, respectively.  
For the fits in time intervals, I, II  and IV, the XRT, BAT, and LAT data are fitted with $C_{\rm stat}$, $\chi^2$, and $PG_{\rm stat}$ statistics, respectively, and 
$S$ is the sum test statistic of $C_{\rm stat}$, $\chi^2$ and $PG_{\rm stat}$.
For the fit in time interval IV, only the XRT data are fitted with $C_{\rm stat}$ statistics, in which case $S$ = $C_{\rm stat}$.

Note: the uncertainties correspond to 90\% confidence level.}
\end{table*}

%\clearpage

\begin{table*}[hb]
\begin{center}
\caption{Fitting result of the spectrum in the brightest part of  during the prompt  emission\label{table:spectrum_prompt}}
\hspace*{-5em} % move the table to left
%\small % added because Table spreads out of a page.
\footnotesize % added because Table spreads out of a page.
\begin{tabular}{cccccccc}
\hline \hline
%\begin{center}

Model  & $\Gamma_{\rm ph,1}$ & $E_{\rm break,1}$ [keV]  & $\Gamma_{\rm ph,2}$ & $E_{\rm break,2}$ [keV] & $\Gamma_{\rm ph,2}$  & $E_{\rm cut}$  [MeV] & $PGstat$/d.o.f.  %\tablenotemark{a}  
\\
%\end{center}
\hline
% total
BKN2  & -0.75$^{+0.04}_{-0.07}$ &  309$^{+130}_{-54}$ & -1.36$^{+0.09}_{-0.21}$ & 2723$^{+2177}_{-699}$ & -2.95$^{+0.12}_{-0.19}$   & -- &  831/577   \\
BKN2 + expcut  & -0.72$\pm$0.4 &  270$\pm$47 & -1.12$\pm$0.06 & 1235$^{+264}_{-197}$ & -2.17$^{+0.10}_{-0.11}$   & 25.8$^{+9.1}_{-5.8}$ &  768/576   \\
slow cooling  & -2/3 &  -- & -3/2 &  & -($p$+2)/2   $\sim$ -2.05 & &     \\
fast cooling  & -2/3 &  -- & -($p$+1)/2 $\sim$-1.55 &  & -($p$+2)/2 $\sim$ -2.05   & &     \\
\hline
\end{tabular}
\end{center}
%% Any table notes must follow the \end{tabular} command.
%\tablenotetext{a}{Sample footnote for table~\ref{tbl-2} that wasgenerated with the \LaTeX\ table environment}
%\tablenotetext{b}{Yet another sample footnote for table~\ref{tbl-2}}
%\tablenotetext{c}{Another sample footnote for table~\ref{tbl-2}}
\leavevmode \\{BKN2 represents a smoothly connected broken power-law function with two spectral breaks.
When the electron index of $p \sim$ 2.1 is assumed, either fast or slow cooling scenario of the  synchrotron emission can be acceptable.
\\Note: the uncertainties correspond to 90\% confidence level.}

\end{table*}

%\bibliography{GRB180720B}

\clearpage
% copy the data from output.bbl on overleaf by M.A.

\end{document}